\documentclass[aps, 
10pt,
prl,
twocolumn,
superscriptaddress,
showpacs,preprintnumbers,amsmath,amssymb,
longbibliography,floatfix]{revtex4-2}
\usepackage[
colorlinks=true, 
pdfstartview=FitV, 
linkcolor=blue, 
citecolor=magenta, 
urlcolor=blue, 
bookmarks=true,
bookmarksnumbered=true,
pdftitle={},
pdfauthor={}
]{hyperref}

\usepackage[dvips]{graphicx}
\usepackage{bm,latexsym,amsmath,amssymb,amsfonts,mathrsfs,textcomp}
\usepackage{color}
\input{colordvi.tex}
\usepackage{comment}
\usepackage{simpler-wick}
\usepackage{tikz}
\usepackage[compat=1.1.0]{tikz-feynhand}
\usepackage[normalem]{ulem}

\newcommand\sect[1]{{\it #1.}---}

\renewcommand\Re{\mathop{\mathrm{Re}}}
\renewcommand\Im{\mathop{\mathrm{Im}}}

\newcommand{\diff}{\mathrm{d}}

\newcommand{\pd}[2]{\frac{\partial #1}{\partial #2}}
\newcommand{\dif}[2]{\frac{\diff #1}{\diff #2}}

\newcommand{\with}{\textrm{with}}

\newcommand{\uni}{\textrm{uni}}

\newcommand{\bx}{\bm{x}}

\newcommand{\bzero}{\bm{0}}
\newcommand{\bp}{\bm{p}}

\newcommand{\bk}{\bm{k}}

\newcommand{\calK}{\mathcal{K}}

\newcommand{\tilA}{\tilde{A}}

\newcommand{\dis}{\displaystyle}

\definecolor{markus}{rgb}{0.09,0.65,0.63}

\definecolor{niko}{rgb}{0.3176,0.612,0.541}

\begin{document}

\title{
Stable-fixed-point description of square pattern formation in driven two-dimensional Bose-Einstein condensates
}

\author{Keisuke Fujii}
\email{fujii@thphys.uni-heidelberg.de}
\affiliation{Institut f\"{u}r Theoretische Physik, Universit\"{a}t Heidelberg, Philosophenweg 19, 69120 Heidelberg, Germany}

\author{Sarah L. G\"orlitz}
\affiliation{Institut f\"{u}r Theoretische Physik, Universit\"{a}t Heidelberg, Philosophenweg 19, 69120 Heidelberg, Germany}

\author{Nikolas Liebster}
\email{pattern-formation@matterwave.de}
\affiliation{Kirchhoff-Institute f\"{u}r Physik, Universit\"{a}t Heidelberg, Im Neuenheimer Feld 227, 69120 Heidelberg, Germany}

\author{Marius Sparn}
\affiliation{Kirchhoff-Institute f\"{u}r Physik, Universit\"{a}t Heidelberg, Im Neuenheimer Feld 227, 69120 Heidelberg, Germany}

\author{Elinor Kath}
\affiliation{Kirchhoff-Institute f\"{u}r Physik, Universit\"{a}t Heidelberg, Im Neuenheimer Feld 227, 69120 Heidelberg, Germany}

\author{Helmut Strobel}
\affiliation{Kirchhoff-Institute f\"{u}r Physik, Universit\"{a}t Heidelberg, Im Neuenheimer Feld 227, 69120 Heidelberg, Germany}

\author{Markus K. Oberthaler}
\affiliation{Kirchhoff-Institute f\"{u}r Physik, Universit\"{a}t Heidelberg, Im Neuenheimer Feld 227, 69120 Heidelberg, Germany}

\author{Tilman Enss}
\affiliation{Institut f\"{u}r Theoretische Physik, Universit\"{a}t Heidelberg, Philosophenweg 19, 69120 Heidelberg, Germany}

\begin{abstract}
We investigate pattern formation in two-dimensional Bose-Einstein condensates (BECs) caused by periodic driving of the interatomic interaction.
We show that this modulation generically leads to a stable square grid density pattern, due to nonlinear effects beyond the initial Faraday instability.
We take the amplitudes of two waves parametrizing the two-dimensional density pattern as order parameters in pattern formation. For these amplitudes, we derive a set of coupled time evolution equations from the Gross--Pitaevskii (GP) equation with a time-periodic interaction.
We identify the fixed points of the time evolution and show by stability analysis
that the inhomogeneous density exhibits a square grid pattern, which can be understood as a manifestation of a stable fixed point.
Our stability analysis establishes the pattern in BECs as a nonequilibrium steady state.
\end{abstract}

\maketitle

\sect{Introduction}%
In recent years, there has been intense interest in many-body dynamics far from equilibrium.
Among these dynamics, pattern formation is particularly intriguing, as uniform states spontaneously become inhomogeneous when external parameters change.
As a spontaneous breaking of symmetry in dynamics, pattern formation appears in nature at diverse scales, not only in physics~\cite{Cross:1993,Cross:2009} but also in chemical reactions~\cite{turing1952} and biology~\cite{Koch:1994}.
However, important questions remain unanswered due to the need for highly controlled experiments.
Thus, ultracold atomic systems are well suited to study pattern formation, due to their advanced experimental techniques.

In BECs, spontaneous pattern formation arises from temporal modulations of system parameters, such as the magnitudes of trapping potentials and interactions~\cite{Staliunas2002}.
This parametric instability is also known as the Faraday instability, as an analogy to a similar phenomenon in classical fluids~\cite{Faraday:1831,Milner:1991,Zhang:1996,Wagner:2000,Simula:2023}.
Indeed, observations in BECs include one-dimensional patterns~\cite{Engels:2007,Nguyen:2019,Hernandez:2021} and surface patterns of a two-dimensional system~\cite{Kwon:2021}.
Furthermore, in two-dimensional systems, the symmetry of selected patterns can be engineered through multiple simultaneous modulations of the atomic interactions~\cite{Zhang:2020}.
Theoretically, parametric instabilities derived from linear analysis in driven one-dimensional quantum gases have been actively studied, and 
many studies have calculated the wavenumber of excitations generated by the instability%
~\cite{
Staliunas:2004,Kevrekidis:2004,Nicolin:2007,
Bhattacherjee:2008,Capuzzi:2008,Nicolin:2010,
Capuzzi:2011,Tang:2011,Nicolin:2011,Sabbatini:2011,
Balaz:2012,Nicolin:2012,Lakomy:2012,
Abdullaev:2013,Balaz:2014,Cairncross:2014,Abdullaev:2015,
Conforti:2016,Sudharsan:2016,Tomio:2017,
Zhu:2019,Vudragovic:2019,Abdullaev:2019,Turmanov:2020,Maity:2020,
Okazaki:2021,Bougas:2021,Otlaadisa:2021,Cheng:2021,Diaz:2021,
Zhang:2022,Shukuno:2023}.

However, previous studies have neglected the effects of nonlinearities and their role in stabilizing certain geometries of patterns far beyond the linear regime. 
This is particularly relevant in two dimensions, where structures are characterized by both lattice symmetry and wavenumber. 
In our recent companion work~\cite{Nikolas:2023}, we observed two-dimensional stable patterns, clearly exhibiting a square grid, with a single-frequency modulation of the interaction without fine-tuning.
In particular, this emergence of square grids contrasts with previous theoretical claims of an oblique grid realization in Ref.~\cite{Staliunas2002}, underlining the importance of a theoretical investigation of the underlying mechanisms.

\begin{figure}[t]
 \centering
 \includegraphics[width=0.93\linewidth]{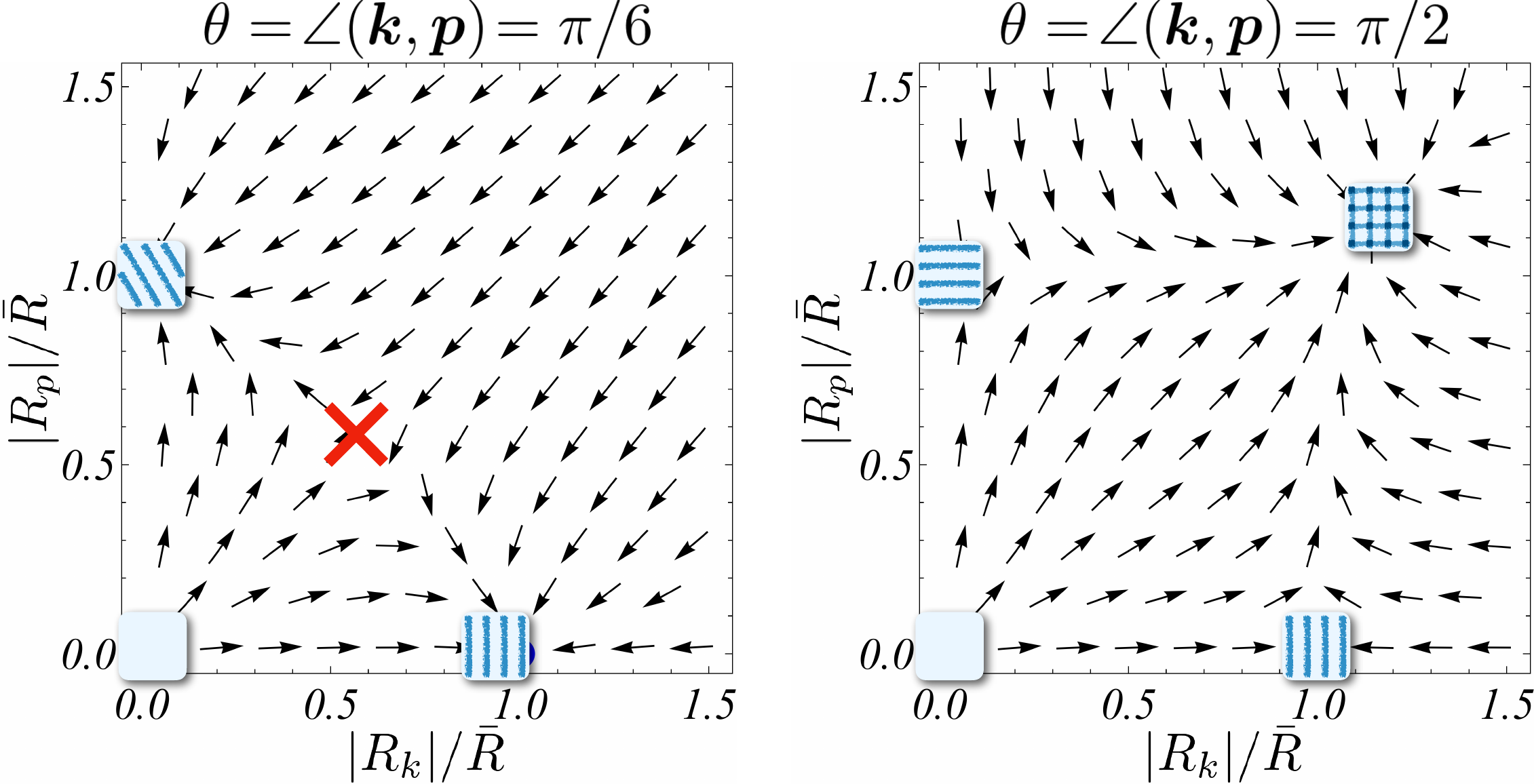}
 \caption{Global stability of the patterns formed by two standing waves in a planar BEC in directions $\bk$ and $\bp$.
Schematic figures at the fixed points represent corresponding stationary solutions, i.e., the grid-pattern, stripe-pattern, and uniform solutions. 
The angle $\theta$ between the two excited modes is $\pi/6$ (left) and $\pi/2$ (right). In the latter case, the square grid pattern emerges as a stable fixed point.
The parameters are $\omega/\mu=2$ and $A=0.6$ with low dissipation $\Gamma=0.1\alpha$, where $\omega,\mu$ and $\alpha$ are the driving frequency, the chemical potential, and the drive amplitude, respectively.} 
 \label{fig:flow}
\end{figure}

In this Letter, we introduce an analytical method that describes not only the growth but also the stabilization of certain patterns due to nonlinearities in driven, single-component BECs.
We derive time-evolution equations for the amplitudes of density waves spanning two-dimensional patterns in BECs. We show that driven condensates exhibit a square grid as a stable pattern.
In our model, we consider density modulations in two directions $\bk$ and $\bp$ in the plane with amplitudes $R_{k}$ and $R_{p}$, respectively. When the angle between the two directions is near $\pi/2$, both amplitudes grow to finite values and we find that the system exhibits a stable grid pattern.
Conversely, for small angles, only one of the amplitudes grows to a finite value while the other is suppressed, and the BEC exhibits a stripe pattern.
This result is clearly seen in Fig.~\ref{fig:flow}, where the global stability of the patterns differs significantly depending on the angle between the two directions.
In the experiment, many modes at different angles initially grow due to the instability, and our analysis shows that two of these modes at an angle close to $\pi/2$ will reinforce each other and grow into a grid pattern, while other modes at small angles are suppressed.

\sect{BEC with a time-periodically modulated interaction}%
\label{sec:GPequation}%
We consider a two-dimensional BEC with an interaction strength $g(t)=\bar{g}[1-A\sin(\omega t)]$ with drive amplitude $|A|<1$ around its mean value $\bar{g}$.
The dynamics of a BEC with wave function $\Psi(t,\bx)$ is described by the GP equation~\cite{Pethick2008}:
\begin{equation}
i\pd{}{t}\Psi(t,\bx)
=\biggl[
	-\frac{\nabla^{2}}{2m}+g(t)|\Psi(t,\bx)|^{2}
\biggr]\Psi(t,\bx),
\label{eq:GPequation}
\end{equation}
where we set $\hbar=1$. We assume an infinitely extended BEC without trapping potential term.

Equation~\eqref{eq:GPequation} has a uniform solution, $\Psi_{\uni}(t)=\Psi_{0}\exp[-i\mu t-i(\mu/\omega)A\cos(\omega t)]$ with the chemical potential $\mu=\bar{g}|\Psi_{0}|^{2}$, but this solution becomes unstable due to the Faraday instability induced by the oscillating interaction~\cite{Staliunas2002}.
This instability can be understood to be caused by a selective amplification of excited modes with wavevector $\bk$ satisfying the resonance condition $n\omega/2=E_{\bk}$ for $n\in\mathbb{N}$.
Here, $E_{\bk}=\sqrt{\epsilon_{\bk}(\epsilon_{\bk}+2\mu)}$ and $\epsilon_{\bk}=\bk^{2}/(2m)$ represent the Bogoliubov quasiparticle and single-particle dispersions, respectively.
The resonance condition $n\omega/2=E_{\bk}$ comes from the fact that the energy quantum $n\omega$, injected into the system by the oscillation, excites two Bogoliubov modes characterized by wavevectors $\bk$ and $-\bk$. 
Within a linear stability analysis, one can indeed derive Mathieu's differential equation from Eq.~\eqref{eq:GPequation}, which shows the amplification of modes with wavenumbers around the resonance condition~\cite{Landau-Lifshitz:mechanics}.

\sect{The amplitude equation}%
\label{sec:amplitude-eq}%
Beyond this linear instability analysis, large occupations and non-zero background interactions can lead to competition between the exponential growth and nonlinear suppression of growth.
Additionally, in two dimensions, non-parallel density waves are coupled due to this nonlinearity, leading to stable grid patterns.
We determine the magnitude of the amplitude and the angle of the realized grid, assuming that the drive amplitude $A$ is small.
In this case, the amplitude of the density pattern grows slowly and its time evolution is systematically obtained as slow-timescale dynamics.
Using the multiple-scale method~\cite{Cross:1993,Cross:2009}, we derive the time-evolution equation for the pattern amplitude from Eq.~\eqref{eq:GPequation}.

We expand the wave function as~\footnote{As discussed in the first paragraph of the Discussion, there is no relevant scattering between three modes satisfying $\bk_1+\bk_2+\bk_3=\bzero$ in our system due to the constraints imposed by the rotating-wave approximation, and scattering between four modes with $(\bk,-\bk,\bk,-\bk)$ and $(\bk,-\bk,\bp,-\bp)$ is dominant. For this reason, the ansatz with two distinct modes, as in Eq.~\eqref{eq:expansion-Psi}, is sufficient.}
\begin{equation}
\Psi(t,\bx)
 = \Psi_{\uni}(t)\biggl[
	1
	+ \phi_{k}(t)\cos(\bk\cdot\bx)
	+ \phi_{p}(t)\cos(\bp\cdot\bx)
\biggr].
\label{eq:expansion-Psi}
\end{equation}
At small drive amplitude $A$, the excitation $\phi_{k/p}$ is naturally expressed in the Bogoliubov basis:
\begin{align}
\phi_{k/p}(t)
&= \biggl(1-\frac{\epsilon_{\bk/\bp}+2\mu}{E_{\bk/\bp}}\biggr)R_{k/p}(t)e^{i\omega t/2} \nonumber \\
&\quad
+ \biggl(1+\frac{\epsilon_{\bk/\bp}+2\mu}{E_{\bk/\bp}}\biggr)R^{\ast}_{k/p}(t)e^{-i\omega t/2},
\label{eq:Bogoliubov-basis}
\end{align}
where the complex amplitudes $R_{k/p}(t)$ obey a complex Ginzburg-Landau type equation (for the detailed derivation, see the Supplement Material~\cite{suppl}):
\begin{align}
&i\dif{}{t}R_{k}(t)
=
\Delta R_{k}(t)
- i\alpha R_{k}^{\ast}(t)
+ \lambda\Bigl(
	|R_{k}(t)|^{2}R_{k}(t) \nonumber \\
&\quad
	+ c_{1}(\theta)|R_{p}(t)|^{2}R_{k}(t)
	+ c_{2}(\theta)R_{p}(t)^{2}R^{\ast}_{k}(t)
\Bigr), \label{eq:amp-eq}
\end{align}
with detuning $\Delta=\omega/2-E_{\bk}$, drive amplitude for the Bogoliubov mode $\alpha=\mu A \epsilon_{\bk}/(2E_{\bk})$, and nonlinearity $\lambda=\mu(5\epsilon_{\bk}+3\mu)/E_{\bk}$.
The same equation holds for $R_{p}(t)$ after exchange of the $\bk$ and $\bp$ labels.
In order to focus on the angle of the realized pattern, we assume that the absolute values of the wavevectors $\bk$ and $\bp$ are equal, as determined by the resonance condition $\Delta=0$ for $n=1$, and set $\epsilon_{\bk}=\epsilon_{\bp}=\epsilon$ and $E_{\bk}=E_{\bp}=E$.
The coupling coefficients $c_{1}(\theta)$ and $c_{2}(\theta)$ between modes in different directions are then given as functions of the angle $\theta \in [0,\pi/2]$ between $\bk$ and $\bp$,
\begin{subequations}
\begin{align}
&c_{1}(\theta)
= \frac{\mu}{5\epsilon+3\mu}
\biggl[
	4\frac{\epsilon^{2}-\mu^{2}}{\mu\epsilon}
	+\biggl(
		\frac{2\epsilon+\mu}{\epsilon}\frac{2\epsilon+\mu}{\epsilon_{+}/2+\mu} \label{eq:coeff1} \\
&
		-\frac{(2\epsilon-\mu)(\epsilon+2\mu)+(2\epsilon^{2}+\mu^{2})\epsilon_{+}/(2\epsilon)}{E^{2}-E^{2}_{+}/4}
	+ (\epsilon_{+} \rightarrow \epsilon_{-})
	\biggr)
\biggr], \nonumber \\
&c_{2}(\theta)
=\frac{\mu}{5\epsilon+3\mu}
\biggl[
	-2\frac{\epsilon^{2}+3\mu \epsilon+\mu^{2}}{\mu\epsilon} \nonumber \\
&
	+\frac{2\epsilon+\mu}{\epsilon}
	\biggl(
		\frac{2\epsilon+\mu}{\epsilon_{+}/2+\mu}+ (\epsilon_{+}\to \epsilon_{-})
	\biggr)
\biggr], \label{eq:coeff2}%
\end{align}%
\label{eq:coeffs}%
\end{subequations}%
where we introduced $E_{\pm}=\sqrt{\epsilon_{\pm}(\epsilon_{\pm}+2\mu)}$ with $\epsilon_{+}=\epsilon_{\bk+\bp}=4\epsilon\cos^{2}\frac{\theta}{2}$ and $\epsilon_{-}=\epsilon_{\bk-\bp}=4\epsilon\sin^{2}\frac{\theta}{2}$~%
\footnote{Since we have introduced the wavevectors $\bk$ and $\bp$ via the cosines in Eq.~\eqref{eq:expansion-Psi}, the signs of them are irrelevant.
Thus, we can take the angle $\theta$ being in $[0,\pi/2]$, and the case with $\theta\in [\pi/2,\pi]$ is given by replacing $\theta\in[0,\pi/2]$ with $\pi-\theta$.}.

Equation~\eqref{eq:amp-eq} is referred to as the amplitude equation. While also other pattern-forming phenomena in classical liquids are understood from amplitude equations, these equations cannot be applied to BECs
because incompressibility is typically assumed for classical fluids,
e.g., for the case of Faraday patterns in water~\cite{Milner:1991,Zhang:1996}.
Instead, BECs are compressible and exhibit density and phase fluctuations, which are reflected in the complex-valued nature of the amplitudes.
Accordingly, our amplitude equation has a nonlinear term proportional to $c_2(\theta)$, which is absent in other amplitude equations for standing-wave patterns.
This distinction highlights the differences in the pattern stabilization mechanism between BECs and classical incompressible fluids.

\sect{The fixed points and their stability analysis}%
\label{sec:fixed-points}%
The time-dependent solutions of the amplitude equation trace out trajectories in the four-dimensional space of the two complex amplitudes. In the following, we analyze their fixed points and stability.
We focus only on the populated modes satisfying the resonance condition at zero detuning $\Delta=0$, and introduce dissipation $\Gamma>0$ phenomenologically to capture, for instance, the suppression from interaction with other modes besides the $\bk$- and $\bp$-modes:~%
\begin{align}
&i\dif{}{t}R_{k}(t)
=
-i\Gamma R_{k}(t)
-i\alpha R_{k}^{\ast}(t)
+ \lambda\Bigl(
	|R_{k}(t)|^{2}R_{k}(t) \nonumber \\
&\quad
	+ c_{1}(\theta)|R_{p}(t)|^{2}R_{k}(t)
	+ c_{2}(\theta)R_{p}(t)^{2}R^{\ast}_{k}(t)
\Bigr).\label{eq:amp-eq-with-D}
\end{align}
The dissipation term effectively captures dynamics beyond the two modes of interest, leading to irreversible evolution of the reduced system, even though the underlying GP equation is reversible~\footnote{As will be discussed below, the magnitude of $\Gamma$ is irrelevant as long as the parameters allow for the occurrence of the Faraday instability at $\alpha>\Gamma$.}.
Setting $\diff R_{k}(t)/\diff t=0$ in Eq.~\eqref{eq:amp-eq-with-D} and similarly for $R_p(t)$, we find four possible fixed-point values of $R_{k}$ and $R_{p}$:
\begin{subequations}
\begin{align}
(\textrm{U})\quad& R_{k}=R_{p}=0, \\
(\textrm{S}_k)\quad&
R_{k}=\bar{R}e^{i\bar{\eta}},\quad
R_{p}=0, \label{eq:fixed-pi-2a}\\
(\textrm{S}_p)\quad& R_{k}=0,\quad R_{p}=\bar{R}e^{i\bar{\eta}}, \\
(\textrm{G})\quad& R_{k}=R_{p}
=\bar{R}e^{i\bar{\eta}}/\sqrt{1+c_{1}(\theta)+c_{2}(\theta)},
\label{eq:grid-amp}
\end{align}
\end{subequations}
with $\bar{R}^{2}=\sqrt{\alpha^{2}-\Gamma^{2}}/\lambda$ and $\exp(i\bar{\eta})=(\sqrt{\alpha-\Gamma}+i\sqrt{\alpha+\Gamma})/\sqrt{2\alpha}$.
The fixed points correspond to the following density patterns: (U) a uniform pattern, (S$_k$) and (S$_p$)  stripe patterns for each direction, and (G) a grid pattern (see Fig.~\ref{fig:flow}).

We first investigate the stability of the uniform fixed point (U). 
For small $R_{k}(t)$ and $R_{p}(t)$ around (U), 
only the linear terms of Eq.~\eqref{eq:amp-eq-with-D} remain, and the $\bk$- and $\bp$-directions become independent.
By dividing the linearized equation into its real and imaginary parts, we directly find the eigenvalues of the Jacobian (scaling dimensions) as $\Lambda=-\Gamma-\alpha,-\Gamma+\alpha$.
Since the amplitudes scale as $e^{\Lambda t}$ around the fixed point (unstable for $\Re\Lambda>0$, stable for $\Re\Lambda<0$, and center for $\Re\Lambda=0$), the fixed point (U) is unstable for $\alpha>\Gamma$.
This corresponds to the Faraday instability, in which the uniform solution becomes unstable when the drive is stronger than the dissipation.

\begin{figure}[t]
 \centering
 \includegraphics[width=0.84\linewidth]{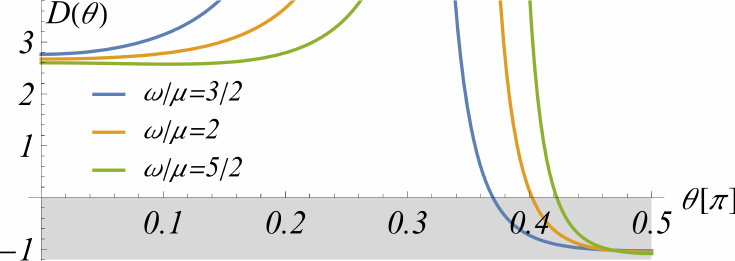}
 \caption{\label{fig:stability-condition}
Stability criterion for the grid pattern:~$D(\theta)$ as a function of $\theta=\angle(\bk,\bp)\in [0,\pi/2]$ with fixed $A=0.6$ for different values of $\omega/\mu$.
The grid pattern for angles $\theta\approx\pi/2$ is stable where $D(\theta)<0$ (gray-shaded region).}
\end{figure}

We next study the stability of the grid fixed point (G).
The four eigenvalues of the amplitude equation linearized around (G) are found to be
\begin{subequations}
\begin{align}
&\Lambda_{1}^{\pm}=-\Gamma\pm i\sqrt{4\alpha^{2}-5\Gamma^{2}}, \\
&\Lambda_{2}^{\pm}=-\Gamma
\pm\frac{\sqrt{4(\alpha^{2}-\Gamma^{2})D(\theta)+\Gamma^{2}(1+c_{1}+c_{2})^{2}}}{1+c_{1}+c_{2}}
\end{align}\label{eq:eigenvalues}
\end{subequations}
with $D(\theta)\equiv -1+c_{1}(\theta)^{2}+2c_{2}(\theta)-c_{2}(\theta)^{2}$.
While the real part of the first eigenvalue $\Lambda^{\pm}_{1}$ is negative for $\alpha>\Gamma$, the second eigenvalue $\Lambda_{2}$ has a negative real part only when
\begin{equation}
D(\theta)
<0. \label{eq:stability-condition}
\end{equation}
The inequality~\eqref{eq:stability-condition} thus provides the condition for the grid pattern to be stable, regardless of the magnitude of the dissipation $\Gamma$.
Note that the stability analysis around the fixed points (S$_k$) and (S$_p$) leads to the inverse inequality $D(\theta)>0$, indicating that stripes are stable when grids are unstable and vice versa.
As seen in Fig.~\ref{fig:stability-condition}, the grid pattern is stable around an angle of $\theta=\pi/2$, which is consistent with the experimental results~\cite{Nikolas:2023}.
Note that the coefficient $c_2(\theta)$ is negative around $\theta=\pi/2$ and therefore enhances the stability of the square pattern.

In the absence of dissipation, the eigenvalues always appear in positive and negative pairs, such as Eq.~\eqref{eq:eigenvalues} with $\Gamma=0$, because the amplitude equation without dissipation fulfills the time-reversal symmetry inherited from the GP equation.
The behavior of the solution in the four-dimensional space near the fixed points can be understood separately for each two-dimensional subspace corresponding to each pair of positive and negative eigenvalues.
Without dissipation, the pair of eigenvalues with a real part makes the fixed point in the corresponding two-dimensional subspace a saddle point, while the pure imaginary pair makes the fixed point a center.
A small dissipation $\Gamma<\alpha$ keeps the saddle fixed point as a saddle while turning the center into a stable focus (in-spiral).

Let us investigate the global behavior of the solutions of the amplitude equation beyond the local behavior around the fixed points.
When we introduce the real and imaginary parts of the phase-rotated amplitudes as $\rho_{k/p}=\Re[R_{k/p}e^{-i\bar{\eta}}]$ and $\nu_{k/p}=\Im[R_{k/p}e^{-i\bar{\eta}}]$, all four fixed points lie in the two-dimensional subspace spanned by $\rho_k$ and $\rho_p$ with $\nu_{k}=\nu_{p}=0$.
The four-dimensional flow trajectories still depart from an unstable fixed point (saddle) and approach an attractive one (in-spiral).
This global behavior can be visualized by utilizing the square eigenvalue $\Lambda^2$, which is positive (repulsive) for real $\Lambda$ at the saddle, while it is negative (attractive) for imaginary $\Lambda$ at the in-spiral.
We can efficiently obtain the square eigenvalues via the second-order differential equation derived from the amplitude equation~\eqref{eq:amp-eq-with-D}, which is given by
\begin{align}
&\dif{^{2}}{t^{2}}\rho_{k}(t)\biggl.\biggr|_{\nu_{k}=\nu_{p}=0}
=\lambda^{2} \biggl[
    \bar{R}^{2}
    + \rho_{k}(t)^{2}
    + (c_{1}-c_{2})\rho_{p}(t)^{2}
\biggr] \nonumber \\
&\times
\biggl[\bar{R}^{2} -\rho_{k}(t)^{2}-(c_{1}+c_{2})\rho_{p}(t)^{2}\biggr]\rho_{k}(t) \label{eq:second-order-diff} \\
&
+2\lambda^{2} c_{2}\biggl[\bar{R}^{2} -\rho_{p}(t)^{2}-(c_{1}+c_{2})\rho_{k}(t)^{2}\biggr]\rho_{p}(t)^{2}\rho_{k}(t) \nonumber
\end{align}
and likewise for $\rho_p(t)$ after exchanging the $\rho_{k}(t)$ and $\rho_{p}(t)$ variables.
The force field described by the right-hand side of Eq.~\eqref{eq:second-order-diff} captures the global behavior of the solution of the original amplitude equation.
This global behavior is shown in Fig.~\ref{fig:flow}, and it changes drastically depending on the angle between $\bk$ and $\bp$~%
\footnote{Figure~\ref{fig:flow} shows the direction of the second derivative given by Eq.~\eqref{eq:second-order-diff}, which represents the force.
Also, since $\rho_{k/p}$ is equal to the absolute value of the amplitude $R_{k/p}$ on the plane with fixed $\nu_{k}=\nu_{p}=0$, we simply denote $\rho_{k/p}$ by $|R_{k/p}|$.}.

\sect{The three-mode amplitude equation}%
\label{sec:complete-theory}%
At a specific angle satisfying $2E=E_{+}$, two Bogoliubov modes with wavevectors $\bk$ and $\bp$ can resonantly scatter into a Bogoliubov mode with wavevector $\bk+\bp$ without violating energy conservation, which enhances the contribution of this collision process. Therefore, the amplitude of the wavevector $\bk+\bp$, which grows proportionally to both the amplitudes $R_{k}$ and $R_{p}$, cannot be neglected, and its omission in the two-mode ansatz~\eqref{eq:expansion-Psi} causes a divergence of the coefficient $c_{1}(\theta)$ in Eq.~\eqref{eq:coeff1} at this specific angle. This same divergence is seen in Fig.~\ref{fig:stability-condition}, e.g., at $\theta\approx 0.34\pi$ for $\omega/\mu=2$.
We now present the complete theory without divergence and show that no other (e.g., triangular) patterns appear near this singular angle.
By additionally including the $\bk+\bp$ mode described by the complex amplitude $R_+(t)$ in the ansatz~\eqref{eq:expansion-Psi}, we derive the coupled amplitude equations for the three modes $R_{k/p}(t)$ and $R_{+}(t)$ as (for details of the derivation and the coefficients see~\cite{suppl})
\begin{subequations}
\begin{align}
&i\dif{}{t}R_{k}(t)
=
-i\Gamma R_{k}(t)
- i\alpha R_{k}^{\ast}(t)
- \beta(\theta)R_{+}(t)R_{p}^{\ast}(t) \nonumber \\
&
+ \lambda\Bigl(
	|R_{k}(t)|^{2}R_{k}(t)
	+ \tilde{c}_{1}(\theta)|R_{p}(t)|^{2}R_{k}(t) \nonumber \\
&
	+ c_{2}(\theta)R_{p}(t)^{2}R^{\ast}_{k}(t)
\Bigr), \label{eq:amp-eq-with-Rkp-1} \\
&i\dif{}{t}R_{+}(t)
=
- i\Gamma_{+}R_{+}(t)
+ \Delta_{+}(\theta) R_{+}(t) \nonumber \\
&
- \beta_{+}(\theta)R_{k}(t)R_{p}(t)
+ \lambda_{+}(\theta) |R_{+}(t)|^{2}R_{+}(t), \label{eq:amp-eq-with-Rkp-2}%
\end{align}%
\label{eq:amp-eq-with-Rkp}%
\end{subequations}%
where $\Gamma$ and $\Gamma_{+}$ are dissipation coefficients.
We find that the coefficient $\tilde{c}_{1}(\theta)$ has no divergence at the singular angle satisfying $2E=E_{+}$ and all coefficients in Eq.~\eqref{eq:amp-eq-with-Rkp} are regular in $\theta\in [0,\pi/2]$.
We note that the $\bk$ and $\bp$ modes now interact with the $\bk+\bp$ mode via quadratic terms in the amplitude equation (three-mode scattering).

\begin{figure}[t]
 \centering
 \includegraphics[width=0.94\linewidth]{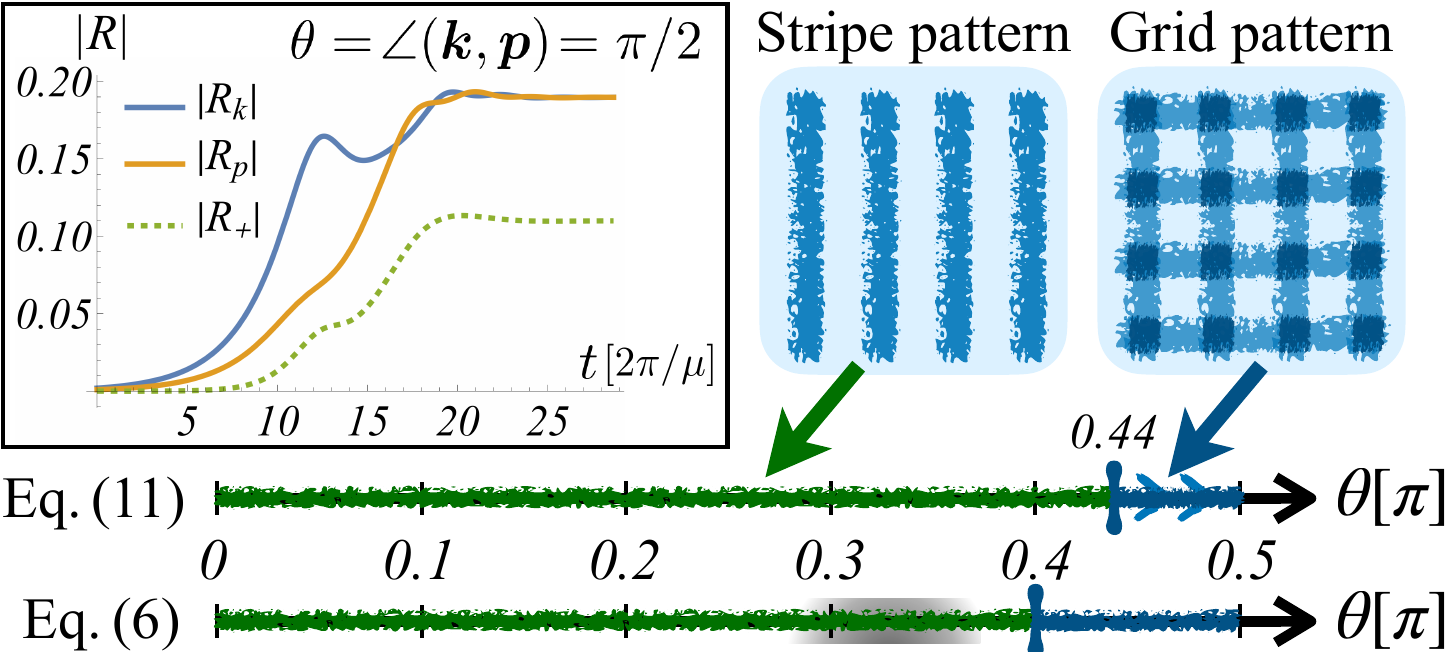}
 \caption{\label{fig:diagram}
Phase diagram of the stable patterns in the three-mode model~\eqref{eq:amp-eq-with-Rkp} and the two-mode model~\eqref{eq:amp-eq-with-D} with parameters $(A,\omega/\mu,\Gamma/\alpha,\Gamma_+/\alpha)=(0.6,2,0.5,1)$.
While angles in the green region have stable stripe solutions, angles in the blue region have stable lattice solutions, as shown schematically in the top right corner.
The phase diagram for Eq.~\eqref{eq:amp-eq-with-D} is obtained analytically from the inequality~\eqref{eq:stability-condition} and corresponds to Fig.~\ref{fig:stability-condition}.
In both cases, the grid pattern is stable around $\theta=\pi/2$, whereas the stripe pattern appears at small angles.
Thus, both models predict the same grid pattern phase, although the two-mode model is incomplete around the singular angle $\theta\approx 0.34\pi$ (gray-shaded region).
Arrows in the phase diagram for Eq.~\eqref{eq:amp-eq-with-Rkp} indicate the stability to changes in the grid angle $\theta$, with $\theta=\pi/2$ being the most stable as discussed in the Supplemental Material~\cite{suppl}.
The top left inset shows the numerical simulation of Eq.~\eqref{eq:amp-eq-with-Rkp} at $\theta=\pi/2$ with initial condition $(R_k,R_p,R_+)=(0.002i,0.001i,0.0001)$, where initial stripes after approximately 10 cycles develop into a grid pattern after approximately 16 cycles.}
\end{figure}

Using Eq.~\eqref{eq:amp-eq-with-Rkp}, we can numerically simulate the time evolution for various initial conditions at a fixed angle between $\bk$ and $\bp$.
The top left part of Fig.~\ref{fig:diagram} shows the simulation results of the time evolution of the amplitudes from an initial state close to a uniform pattern at $\theta=\pi/2$, demonstrating that the system arrives at a grid pattern with all amplitudes finite.
Repeating these simulations for different angles $\theta$, we arrive at the phase diagrams for the three-mode and two-mode models shown in Fig.~\ref{fig:diagram}.
In both phase diagrams, the grid pattern is stable in a region around $\theta=\pi/2$, and the phase diagrams are in good agreement.
Notably, the three-mode model~\eqref{eq:amp-eq-with-Rkp} has no singularity in its coefficients and its phase diagram is reliable over the entire range of angles $\theta\in[0,\pi/2]$.
Furthermore, using Eq.~\eqref{eq:amp-eq-with-Rkp} we investigate the stability to changes in the grid angle $\theta$ and numerically confirm that the square grid with $\theta=\pi/2$ is the most stable among the stable grids as indicated by the arrows in the phase diagram of Fig.~\ref{fig:diagram}
(see the Supplemental Material~\cite{suppl} for a detailed discussion).

\sect{Discussion and outlook}%
\label{sec:Discussion}%
Before concluding, we briefly discuss the scattering between the three modes satisfying $\bk_1\pm\bk_2\pm\bk_3=\bzero$, which usually leads to a triangular pattern~\cite{Cross:2009}.
In the Faraday pattern formation, the leading three-mode scattering occurs at the frequency $\omega/2\pm\omega/2\pm\omega/2$ because the excited modes have energy $\omega/2$ from the first resonance condition $n=1$.
Because $\omega/2\pm\omega/2\pm\omega/2\neq 0$, the three-mode scattering is a fast-rotating contribution in the rotating wave basis and can be neglected.
Instead, if the third mode satisfies the second resonance condition $n=2$ with frequency $\omega$ instead of $\omega/2$, three-mode scattering can become relevant since it has a slow-rotating contribution with $\omega/2 +\omega/2-\omega=0$.
In fact, the amplitude equation~\eqref{eq:amp-eq-with-Rkp} includes this three-mode scattering.
However, we found that the stability of the patterns does not change significantly between Eqs.~\eqref{eq:amp-eq} and~\eqref{eq:amp-eq-with-Rkp}.

In this Letter, we have derived the amplitude equations~\eqref{eq:amp-eq-with-D} and \eqref{eq:amp-eq-with-Rkp} for pattern formation in two-dimensional BECs beyond the Faraday instability.
The amplitude equation can be considered as a complex Ginzburg-Landau type equation for pattern formation with the amplitudes in the two directions as order parameters, and it provides a simple description of the system dynamics.
Our method to derive the amplitude equation is equivalent to the renormalization-group theory for asymptotic analysis~\cite{Chen1994,Kunihiro1995,Chen1996}.
Accordingly, the amplitude equation describes the order parameter dynamics as an effective model for the only two or three relevant modes remaining at long times, while in a renormalization group sense it incorporates the multitude of irrelevant modes of the full GP simulation in the dissipation coefficient.

Using the obtained amplitude equation, we have analyzed the stability between the uniform, stripe-pattern, and grid-pattern solutions.
For $\alpha>\Gamma$, where the drive amplitude is stronger than the dissipation, the uniform solution becomes unstable, resulting in an inhomogeneous density pattern.
Figures~\ref{fig:stability-condition} and \ref{fig:diagram} show that the grid pattern becomes stable around the angle $\pi/2$ between the two excitation directions.
The global stability of the amplitude is shown in Fig.~\ref{fig:flow}.
In particular, in the three-mode model, taking into account the resonant scattering that occurs around the singular angles, we have shown that the grid is stable at $\pi/2$ and not at the singular angle previously suggested.
Our results provide the theoretical framework for the experimental data presented in our companion paper~\cite{Nikolas:2023}.
Furthermore, the amplitude equation has been experimentally validated under various initial conditions and has been confirmed to give a good description.
Our stability analysis has established patterns in superfluids as nonequilibrium steady states.
Moreover, these patterns fit the definition of supersolidity as a self-stabilized superfluid state with spontaneous translational symmetry breaking, suggesting the emergence of a new type of supersolidity as a nonequilibrium steady state; we leave the search for this novel quantum phase to future work.

Another interesting direction for future research is the further study of amplitude equations in BECs.
By their complex-valued nature, they are of a new type not seen in other instances of pattern formation, and it will be crucial to understand their properties in applications to two-dimensional dipolar or spinor BECs~\cite{Nath:2010,Chen:2019,Jose:2023}.
Furthermore, BECs have been established as field simulators, e.g., in gravity analogs, and the amplitude equation could provide significant insight into instabilities and their stabilization mechanisms beyond the realm of pattern formation.

\smallskip
\begin{acknowledgments}
The authors thank T.~Simula and F.~Ziebert for stimulating discussions.
This work was supported by Deutsche Forschungsgemeinschaft (German Research Foundation) under Project No.~273811115 (SFB1225 ISOQUANT) and under Germany's Excellence Strategy EXC2181/1-390900948 (the Heidelberg STRUCTURES Excellence Cluster).
This project was funded within the QuantERA II Programme that has received funding from the European Union's Horizon 2020 research and innovation program under Grant Agreement No.~101017733.
N.L. acknowledges support from the Studienstiftung des Deutschen Volkes.
\end{acknowledgments}

\bibliography{faraday-pattern}

\clearpage
\appendix
\pagebreak
\widetext
\begin{center}
\textbf{\large Supplemental Materials:\\
Stable Fixed Point Description of Square Pattern Formation in Driven Two-Dimensional Bose-Einstein Condensates
}
\end{center}
\setcounter{equation}{0}
\setcounter{figure}{0}
\setcounter{table}{0}
\setcounter{page}{1}
\newcounter{supplementeqcountar}
\newcommand\suppsect[1]{{\it #1.}---}
\renewcommand{\theequation}{S\arabic{equation}}
\renewcommand{\thefigure}{S\arabic{figure}}

\section{Derivation of the amplitude equation~\eqref{eq:amp-eq}}
\label{sec:derivation}
We perform a perturbative expansion of the wave function $\Psi(t,\bx)$ obeying the GP equation,
\begin{align}
i\pd{}{t}\Psi(t,\bx)
=
\biggl[
	-\frac{\nabla^{2}}{2m}+g(t)|\Psi(t,\bx)|^{2}
\biggr]\Psi(t,\bx),
\quad\with\quad
g(t)=\bar{g}[1-A\sin(\omega t)],
\label{suppl-eq:GPequation}
\end{align}
using the multiple-scale method.
This section is partly based on Ref.~\cite{devalcarcel2002}.

\subsubsection{Order counting and perturbative equations}
Introducing a small bookkeeping parameter $\varepsilon$, we denote quantities as
\begin{align}
A=\varepsilon^{2}\tilA,\qquad 
\frac{\omega}{2}-E_{\bk}=\varepsilon^{2}\tilde{\omega}_{k},\qquad 
\frac{\omega}{2}-E_{\bp}=\varepsilon^{2}\tilde{\omega}_{p}.
\end{align}
According to the multiple-scale method, we introduce a slow time $\tau=\varepsilon^{2}t$ and expand the wave function as
\begin{align}
\Psi(t,\bx)
=\Psi_{\uni}(t)
\biggl[
	1
	+ \sum_{n=1}^{\infty}\varepsilon^{n}\phi_{n}(t,\tau,\bx)
\biggr],
\end{align}
where we regarded $\phi_{n}$ as a function not only of $t$ and $\bx$ but also of $\tau$.
Here, $\Psi_{\uni}(t)=\Psi_{0}\exp[-i\mu t-i(\mu/\omega)A\cos(\omega t)]$ is a uniform solution of Eq.~\eqref{suppl-eq:GPequation} with the chemical potential $\mu=\bar{g}|\Psi_{0}|^{2}$.
For later convenience, we expand the wave function around $\Psi_{\uni}(t)$ instead of the zeroth-order solution, $\Psi_{0}e^{-i\mu t}$, of Eq.~\eqref{suppl-eq:GPequation}.
To focus on the two excited modes spanning the two-dimensional pattern, we suppose $\phi_{1}(t,\tau,\bx)$ to have the form of
\begin{align}
\phi_{1}(t,\tau,\bx)=
	\phi^{k}_{1}(t,\tau)\cos(\bk\cdot\bx)
	+\phi^{p}_{1}(t,\tau)\cos(\bp\cdot\bx).
\label{suppl-eq:phi1}
\end{align}

Substituting the above reparameterizations and the expansion, together with $\partial_{t}\to \partial_{t}+\varepsilon^{2}\partial_{\tau}$, into Eq.~\eqref{suppl-eq:GPequation}, we find
\begin{subequations}
\begin{align}
&i\pd{}{t}\phi_{1}(t,\tau,\bx)+\frac{\nabla^{2}}{2m}\phi_{1}(t,\tau,\bx)-2\mu \Re[\phi_{1}(t,\tau,\bx)]=0, 
\label{suppl-eq:1st-eq} \\
&i\pd{}{t}\phi_{2}(t,\tau,\bx)+\frac{\nabla^{2}}{2m}\phi_{2}(t,\tau,\bx)-2\mu \Re[\phi_{2}(t,\tau,\bx)]
=\mu\Bigl(
	|\phi_{1}(t,\tau,\bx)|^{2}
	+2\Re[\phi_{1}(t,\tau,\bx)]\phi_{1}(t,\tau,\bx)
\Bigr), \label{suppl-eq:2nd-eq} \\
&i\pd{}{t}\phi_{3}(t,\tau,\bx)+\frac{\nabla^{2}}{2m}\phi_{3}(t,\tau,\bx)-2\mu \Re[\phi_{3}(t,\tau,\bx)] \nonumber \\
& =
- i\pd{}{\tau}\phi_{1}(t,\tau,\bx)
+ \mu\Bigl(
	-2\tilA\sin(\omega t)\Re[\phi_{1}(t,\tau,\bx)]
	+|\phi_{1}(t,\tau,\bx)|^{2}\phi_{1}(t,\tau,\bx)
	+2\calK(\phi_{1}(t,\tau,\bx),\phi_{2}(t,\tau,\bx))
\Bigr), \label{suppl-eq:3rd-eq}
\end{align}
\end{subequations}
for increasing orders of $\varepsilon$, where we introduced 
\begin{align}
\calK(\phi_{1},\phi_{2})\equiv \Re[\phi_{1}]\phi_{2}
	+\Re[\phi_{2}]\phi_{1}
	+\Re[\phi_{1}\phi^{\ast}_{2}].
\end{align}
The solution of Eq.~\eqref{suppl-eq:1st-eq} describes the usual Bogoliubov quasiparticles and, with the use of Eq.~\eqref{suppl-eq:phi1}, is found as
\begin{align}
\phi_{1}^{k}(t,\tau)
&= \biggl(1-\frac{\epsilon_{\bk}+2\mu}{E_{\bk}}\biggr)r_{k}(\tau)e^{iE_{\bk}t}
+ \biggl(1+\frac{\epsilon_{\bk}+2\mu}{E_{\bk}}\biggr)r^{\ast}_{k}(\tau)e^{-iE_{\bk}t},
\label{suppl-eq:Bogoliubov-basis}
\end{align}
and the same for $\phi^{p}_{1}(t,\tau)$ with $\bk$ replaced by $\bp$.
Our remaining task in the multiple-scale method is to derive the equation determining the complex amplitude $r_{k/p}(\tau)$ from the solvability condition that Eqs.~\eqref{suppl-eq:2nd-eq} and \eqref{suppl-eq:3rd-eq} have no secular terms.

\subsubsection{Solution of the second-order equation}
The general solution of Eq.~\eqref{suppl-eq:2nd-eq} is irrelevant for our purposes; it is sufficient to consider only modes resulting from scattering of the two modes with wavevectors $\bk$ and $\bp$ in $\phi_{1}(t,\tau,\bx)$.
This allow us to take $\phi_{2}(t,\tau,\bx)$ in the form of
\begin{align}
\phi_{2}(t,\tau,\bx)
= \phi_{20}(t,\tau)
+ \phi_{22}^{k}(t,\tau)\cos(2\bk\cdot\bx)
+ \phi_{22}^{p}(t,\tau)\cos(2\bp\cdot\bx)
+ \phi_{2}^{+}(t,\tau)\cos((\bk+\bp)\cdot\bx)
+ \phi_{2}^{-}(t,\tau)\cos((\bk-\bp)\cdot\bx).
\end{align}
Plugging this form into Eq.~\eqref{suppl-eq:2nd-eq} leads to
\begin{align}
i\pd{}{t}\phi_{20}(t,\tau)-2\mu \Re[\phi_{20}(t,\tau)]
=\frac{\mu}{2}\Bigl(
	|\phi^{k}_{1}(t,\tau)|^{2}
	+ |\phi^{p}_{1}(t,\tau)|^{2}
	+2\Re[\phi^{k}_{1}(t,\tau)]\phi^{k}_{1}(t,\tau)
	+2\Re[\phi^{p}_{1}(t,\tau)]\phi^{p}_{1}(t,\tau)
\Bigr) \label{suppl-eq:diff-eq-phi20}
\end{align}
for $\phi_{20}(t,\tau)$,
\begin{align}
i\pd{}{t}\phi^{k}_{22}(t,\tau)-4\epsilon_{\bk}\phi^{k}_{22}(t,\tau)-2\mu \Re[\phi^{k}_{22}(t,\tau)]
=\frac{\mu}{2}\Bigl(
	|\phi^{k}_{1}(t,\tau)|^{2}
	+2\Re[\phi^{k}_{1}(t,\tau)]\phi^{k}_{1}(t,\tau)
\Bigr) \label{suppl-eq:diff-eq-phi22}
\end{align}
for $\phi^{k}_{22}(t,\tau)$ and the same with $\bk$ replaced by $\bp$ for $\phi_{22}^{p}(t,\tau)$, and
\begin{align}
i\pd{}{t}\phi^{+}_{2}(t,\tau)-\epsilon_{+}\phi^{+}_{2}(t,\tau)-2\mu \Re[\phi^{+}_{2}(t,\tau)]
=\mu \calK(\phi^{k}_{1}(t,\tau),\phi^{p}_{1}(t,\tau)) \label{suppl-eq:diff-eq-phi2plus}
\end{align}
for $\phi^{+}_{2}(t,\tau)$ and the same with $\epsilon_{+}=\epsilon_{\bk+\bp}$ replaced by $\epsilon_{-}=\epsilon_{\bk-\bp}$ for $\phi^{-}_{2}(t,\tau)$.

By dividing Eqs.~\eqref{suppl-eq:diff-eq-phi20}, \eqref{suppl-eq:diff-eq-phi22}, and \eqref{suppl-eq:diff-eq-phi2plus} into their real and imaginary parts, and by using Eq.~\eqref{suppl-eq:Bogoliubov-basis}, we can find the solutions to these equations.
First, the solution of Eq.~\eqref{suppl-eq:diff-eq-phi20} is found as
\begin{align}
\phi_{20}(t,\tau)=\phi_{20}^{k}(t,\tau)+\phi^{p}_{20}(t,\tau)
\end{align}
with
\begin{align}
\phi_{20}^{k}(t,\tau)
=-\frac{2\epsilon_{\bk}+\mu}{\epsilon_{\bk}}|r_{k}(\tau)|^{2}
+\frac{\mu}{2\epsilon_{\bk}}\biggl(1-\frac{\epsilon_{\bk}}{E_{\bk}}\biggr)r_{k}(\tau)^{2}e^{2iE_{\bk}t}
+\frac{\mu}{2\epsilon_{\bk}}\biggl(1+\frac{\epsilon_{\bk}}{E_{\bk}}\biggr)r^{\ast}_{k}(\tau)^{2}e^{-2iE_{\bk}t}+i\tilde{v}_{20}^{k}(\tau)
\end{align}
and the same with $\bk$ replaced by $\bp$ for $\phi^{p}_{20}(t,\tau)$.
Here, $\tilde{v}_{20}^{k}(\tau)$ is an arbitrary real function of $\tau$ independent of $t$, which does not contribute to the results.

Second, the solution of Eq.~\eqref{suppl-eq:diff-eq-phi22} is found as
\begin{align}
\phi_{22}^{k}(t,\tau)
=-\frac{\mu}{\epsilon_{\bk}}|r_{k}(\tau)|^{2}
-\frac{\mu}{2\epsilon_{\bk}}\biggl(1-\frac{E_{\bk}}{\epsilon_{\bk}}\biggr)r_{k}(\tau)^{2}e^{2iE_{\bk}t}
-\frac{\mu}{2\epsilon_{\bk}}\biggl(1+\frac{E_{\bk}}{\epsilon_{\bk}}\biggr)r^{\ast}_{k}(\tau)^{2}e^{-2iE_{\bk}t},\label{suppl-eq:phi22k}
\end{align}
and the same with $\bk$ replaced by $\bp$ for $\phi_{22}^{p}(t,\tau)$.
Although $\phi_{22}^{k}(t,\tau)$ has additional terms proportional to $e^{\pm iE_{2\bk}t}$ in addition to the particular solution~\eqref{suppl-eq:phi22k}, we dropped them because those additional terms do not contribute to the results.

Finally, the solution of Eq.~\eqref{suppl-eq:diff-eq-phi2plus} is found in the form of
\begin{equation}
\begin{split}
\Re[\phi_{2}^{+}(t,\tau)]
&= \varphi_{1}^{+} r_{k}(\tau)r_{p}(\tau) e^{i(E_{\bk}+E_{\bp})t}
+ \varphi_{2}^{+} r_{k}(\tau)r^{\ast}_{p}(\tau) e^{i(E_{\bk}-E_{\bp})t}+(\textrm{complex conjugate}), \\
\Im[\phi_{2}^{+}(t,\tau)]
&= i\varphi_{3}^{+} r_{k}(\tau)r_{p}(\tau) e^{i(E_{\bk}+E_{\bp})t}
+ i\varphi_{4}^{+} r_{k}(\tau)r^{\ast}_{p}(\tau) e^{i(E_{\bk}-E_{\bp})t}+(\textrm{complex conjugate}),
\label{suppl-eq:phi2plus}
\end{split}
\end{equation}
where the coefficients $\varphi_{n}^{+}$ for $n=1,2,3,4$ satisfy the following equations and are consequently real:
\begin{align}
\left\{
\begin{array}{l}
\dis (E_{\bk}+E_{\bp})\varphi^{+}_{1}-\epsilon_{+}\varphi^{+}_{3}=\mu\biggl(\frac{E_{\bk}}{\epsilon_{\bk}}+\frac{E_{\bp}}{\epsilon_{\bp}}\biggr), \\[2.0ex]
\dis (E_{\bk}+E_{\bp})\varphi^{+}_{3}-(\epsilon_{+}+2\mu)\varphi^{+}_{1}=\mu\biggl(3-\frac{E_{\bk}}{\epsilon_{\bk}}\frac{E_{\bp}}{\epsilon_{\bp}}\biggr),
\end{array}
\right.
\quad
\left\{
\begin{array}{l}
\dis (E_{\bk}-E_{\bp})\varphi^{+}_{2}-\epsilon_{+}\varphi^{+}_{4}=\mu\biggl(\frac{E_{\bk}}{\epsilon_{\bk}}-\frac{E_{\bp}}{\epsilon_{\bp}}\biggr), \\[2.0ex]
\dis (E_{\bk}-E_{\bp})\varphi^{+}_{4}-(\epsilon_{+}+2\mu)\varphi^{+}_{2}=\mu\biggl(3+\frac{E_{\bk}}{\epsilon_{\bk}}\frac{E_{\bp}}{\epsilon_{\bp}}\biggr).
\end{array}
\right. \label{suppl-eq:coefficients-eq-phi2plus}
\end{align}
We again dropped the additional terms in the solution for $\phi^{+}_{2}(t,\tau)$ because they do not contribute to the results.
The solution of $\phi^{-}_{2}(t,\tau)$ is given in the same form with $\varphi^{+}_{n}$ replaced by $\varphi^{-}_{n}$ in Eq.~\eqref{suppl-eq:phi2plus}, and its coefficients $\varphi^{-}_{n}$ satisfy the same equations~\eqref{suppl-eq:coefficients-eq-phi2plus} with $\varphi^{+}_{n}$ and $\epsilon_{+}$ replaced by $\varphi^{-}_{n}$ and $\epsilon_{-}$, respectively.

\subsubsection{The solvability condition}
Similarly to $\phi_{2}(t,\tau,\bx)$, we expand $\phi_{3}(t,\tau,\bx)$ in the Fourier cosine basis as
\begin{align}
\phi_{3}(t,\tau,\bx)=\phi_{31}^{k}(t,\tau)\cos(\bk\cdot\bx)+\phi_{31}^{p}(t,\tau)\cos(\bp\cdot\bx)+\cdots,
\end{align}
where the ellipsis represents terms with other bases. We need only $\phi_{31}^{k/p}(t,\tau)$ for our purpose.
From Eq.~\eqref{suppl-eq:3rd-eq}, the differential equation for $\phi^{k}_{31}(t,\tau)$ is obtained as
\begin{align}
i\pd{}{t}\phi_{31}^{k}(t,\tau)-\epsilon_{\bk}\phi^{k}_{31}(t,\tau)-2\mu\Re[\phi_{31}^{k}(t,\tau)]=f(t,\tau)
\label{suppl-eq:phi31k-eq}
\end{align}
with
\begin{align}
f(t,\tau)
& =
-i\pd{}{\tau}\phi_{1}^{k}(t,\tau)
+\mu\biggl\{
	-2\tilA \sin(\omega t)\Re[\phi^{k}_{1}(t,\tau)]
	+\frac{3}{4}|\phi^{k}_{1}(t,\tau)|^{2}\phi^{k}_{1}(t,\tau)
	+|\phi^{p}_{1}(t,\tau)|^{2}\phi^{k}_{1}(t,\tau)
	+\frac{1}{2}\phi^{k \ast}_{1}(t,\tau)\phi^{p}_{1}(t,\tau)^{2} \nonumber \\
&\quad
	+ \calK(\phi^{k}_{1}(t,\tau),2\phi_{20}(t,\tau)+\phi^{k}_{22}(t,\tau))
	+ \calK(\phi^{p}_{1}(t,\tau),\phi_{2}^{+}(t,\tau)+\phi^{-}_{2}(t,\tau)) 
\biggr\}.
\end{align}
Here, we write Eq.~\eqref{suppl-eq:phi31k-eq} in the matrix form as
\begin{align}
\pd{}{t}
\left(
\begin{array}{c}
\Re[\phi_{31}^{k}(t,\tau)] \\[1.2ex]
\Im[\phi_{31}^{k}(t,\tau)]
\end{array}
\right)+
\left(
\begin{array}{cc}
0 & -\epsilon_{\bk} \\[1.2ex]
\epsilon_{\bk}+2\mu & 0
\end{array}
\right)
\left(
\begin{array}{c}
\Re[\phi_{31}^{k}(t,\tau)] \\[1.2ex]
\Im[\phi_{31}^{k}(t,\tau)]
\end{array}
\right)
=
\left(
\begin{array}{c}
\Im[f(t,\tau)] \\[1.2ex]
-\Re[f(t,\tau)]
\end{array}
\right).
\end{align}
Diagonalizing this equation, we obtain
\begin{align}
\pd{}{t}\chi(t,\tau)-iE_{\bk}\chi(t,\tau)=g(t,\tau)
\label{suppl-eq:diagonalized-eq}
\end{align}
and its complex conjugate with
\begin{align}
\chi(t,\tau)
&=\frac{1}{\sqrt{2}}\biggl(
	\frac{\epsilon_{\bk}+2\mu}{E_{\bk}}\Re[\phi^{k}_{31}(t,\tau)]
	-i\Im[\phi^{k}_{31}(t,\tau)]
\biggr), \\
g(t,\tau)
&=\frac{1}{\sqrt{2}}\biggl(
	\frac{\epsilon_{\bk}+2\mu}{E_{\bk}}\Im[f(t,\tau)]
	+i\Re[f(t,\tau)]
\biggr).
\end{align}
Therefore, since the homogeneous version of Eq.~\eqref{suppl-eq:diagonalized-eq}, i.e., Eq.~\eqref{suppl-eq:diagonalized-eq} with $g(t,\tau)=0$, has an oscillating solution $e^{iE_{\bk}t}$, the condition for Eq.~\eqref{suppl-eq:diagonalized-eq} to have no secular term is equivalent to that $g(t,\tau)$ has no term proportional to $e^{iE_{\bk}t}$.

Substituting the above results into $g(t,\tau)$ straightforwardly yields
\begin{align}
&\frac{i}{\sqrt{2}}\frac{E_{\bk}}{\epsilon_{\bk}+2\mu}g(t,\tau)
= -i\dif{r_{k}(\tau)}{\tau}e^{iE_{\bk}t}
-i\mu\frac{\tilA}{2}\frac{\epsilon_{\bk}}{E_{\bk}}r^{\ast}_{k}(\tau)e^{2i\tilde{\omega}_{k}\tau}e^{iE_{\bk}t} \nonumber \\
&\quad
+\mu\frac{5\epsilon_{\bk}+3\mu}{E_{\bk}}
\Bigl(
	|r_{k}(\tau)|^{2}r_{k}(\tau)
	+c_{1}(\bk,\bp)|r_{p}(\tau)|^{2}r_{k}(\tau)
	+c_{2}(\bk,\bp)r_{p}(\tau)^{2}r_{k}^{\ast}(\tau)e^{2i(\tilde{\omega}_{k}-\tilde{\omega}_{p})\tau}
\Bigr)e^{iE_{\bk}t}+\cdots
\label{suppl-eq:derivaton}
\end{align}
with
\begin{subequations}
\begin{align}
c_{1}(\bk,\bp)&= 
\frac{E_{\bk}}{5\epsilon_{\bk}+3\mu}
\biggl[
	4\frac{\epsilon_{\bk}\epsilon_{\bp}-\mu^{2}}{\epsilon_{\bp}E_{\bk}}
	-\frac{1}{2}(\varphi^{+}_{1}+\varphi^{-}_{1})
	\biggl(-\frac{\epsilon_{\bp}+2\mu}{E_{\bp}}+3\frac{\epsilon_{\bk}}{E_{\bk}}\biggr)
	-\frac{1}{2}(\varphi^{+}_{2}+\varphi^{-}_{2})
	\biggl(\frac{\epsilon_{\bp}+2\mu}{E_{\bp}}+3\frac{\epsilon_{\bk}}{E_{\bk}}\biggr) \nonumber \\
&\quad
	-\frac{1}{2}(\varphi^{+}_{3}+\varphi^{-}_{3})
	\biggl(1+\frac{\epsilon_{\bp}+2\mu}{E_{\bp}}\frac{\epsilon_{\bk}}{E_{\bk}}\biggr)
	-\frac{1}{2}(\varphi^{+}_{4}+\varphi^{-}_{4})
	\biggl(1-\frac{\epsilon_{\bp}+2\mu}{E_{\bp}}\frac{\epsilon_{\bk}}{E_{\bk}}\biggr)
\biggr], \label{suppl-eq:def-c1-kp}\\
c_{2}(\bk,\bp)
&=
-\frac{E_{\bk}}{5\epsilon_{\bk}+3\mu}
\biggl[
	\frac{\epsilon_{\bk}+\mu}{E_{\bk}}
	+ \frac{\mu}{\epsilon_{\bp}} \frac{2\epsilon_{\bk}+2\mu}{E_{\bk}}
	+ \frac{\epsilon_{\bp}+3\mu}{E_{\bp}} \nonumber \\
&\quad
	+ \frac{1}{2} (\varphi^{+}_{2}+\varphi^{-}_{2})
	\biggl(\frac{\epsilon_{\bp}+2\mu}{E_{\bp}}+3\frac{\epsilon_{\bk}}{E_{\bk}}\biggr)
	- \frac{1}{2} (\varphi^{+}_{4}+\varphi^{-}_{4}) 
	\biggl(1-\frac{\epsilon_{\bp}+2\mu}{E_{\bp}}\frac{\epsilon_{\bk}}{E_{\bk}}\biggr)
\biggr],
\label{suppl-eq:def-c2-kp}
\end{align}
\end{subequations}
where the ellipsis in Eq.~\eqref{suppl-eq:derivaton} represents terms not proportional to $e^{iE_{\bk}t}$.
Note that $e^{iE_{\bp}t}=e^{iE_{\bk}t}e^{i(\tilde{\omega}_{k}-\tilde{\omega}_{p})\tau}$ is considered proportional to $e^{iE_{\bk}t}$ as a function of $t$ because of $E_{\bp}=\frac{\omega}{2}-\varepsilon^{2}\tilde{\omega}_{p}=E_{\bk}+\varepsilon^{2}(\tilde{\omega}_{k}-\tilde{\omega}_{p})$.
Thus, the solvability condition that the coefficient of $e^{iE_{\bk}t}$ in $g(t,\tau)$ is equal to zero is found to be
\begin{align}
i\dif{}{\tau}r_{k}(\tau)& = -i\mu\frac{\tilA}{2}\frac{\epsilon_{\bk}}{E_{\bk}} r^{\ast}_{k}(\tau)e^{2i\tilde{\omega}_{k}\tau}
+\mu\frac{5\epsilon_{\bk}+3\mu}{E_{\bk}}
\Bigl(
	|r_{k}(\tau)|^{2}r_{k}(\tau)
	+c_{1}(\bk,\bp)|r_{p}(\tau)|^{2}r_{k}(\tau)
	+c_{2}(\bk,\bp)r_{p}(\tau)^{2}r_{k}^{\ast}(\tau)e^{2i(\tilde{\omega}_{k}-\tilde{\omega}_{p})\tau}
\Bigr). 
\label{suppl-eq:solvability-condition}
\end{align}

\subsubsection{Resulting amplitude equation}
Assuming the absolute values of $\bk$ and $\bp$ to be equal, we introduce
\begin{align}
\epsilon=\epsilon_{\bk}=\epsilon_{\bp},\qquad
E=E_{\bk}=E_{\bp}=\sqrt{\epsilon(\epsilon+2\mu)}.
\label{suppl-eq:simplification}
\end{align}
Under this assumption, we find the coefficients $\varphi^{+}_{n}$ from Eq.~\eqref{suppl-eq:coefficients-eq-phi2plus} as
\begin{align}
\varphi^{+}_{1}=\frac{\mu\left[\epsilon+2\mu+(\epsilon-\mu)\epsilon_{+}/(2\epsilon)\right]}{E^{2}-E^{2}_{+}/4},\quad
\varphi^{+}_{2}=-\frac{\mu}{\epsilon}\frac{2\epsilon+\mu}{\epsilon_{+}/2+\mu},\quad
\varphi^{+}_{3}=\frac{\mu E\left[1+\epsilon_{+}/(2\epsilon)\right]}{E^{2}-E^{2}_{+}/4},\quad
\varphi^{+}_{4}=0,
\label{suppl-eq:coeff-phi-plus}
\end{align}
where we introduced $E_{+}=\sqrt{\epsilon_{+}(\epsilon_{+}+2\mu)}$.
The coefficients $\varphi_n^{-}$ are obtained by replacing $\epsilon_{+}$ with $\epsilon_{-}$ in Eq.~\eqref{suppl-eq:coeff-phi-plus}.
Under the assumption~\eqref{suppl-eq:simplification}, $\epsilon_{+}=4\epsilon\cos^{2}(\theta/2)$ and $\epsilon_{-}=4\epsilon\sin^{2}(\theta/2)$ turn into functions of the angle between $\bk$ and $\bp$, and thus the coefficients $c_{1}(\bk,\bp)$ and $c_{2}(\bk,\bp)$ are also simplified as functions of the angle,
\begin{subequations}
\begin{align}
c_{1}(\theta)
&=\frac{\mu}{5\epsilon+3\mu}
\biggl[
	4\frac{\epsilon^{2}-\mu^{2}}{\mu\epsilon}
	+\biggl(
		\frac{2\epsilon+\mu}{\epsilon}\frac{2\epsilon+\mu}{\epsilon_{+}/2+\mu}
		-\frac{(2\epsilon-\mu)(\epsilon+2\mu)+(2\epsilon^{2}+\mu^{2})\epsilon_{+}/(2\epsilon)}{E^{2}-E^{2}_{+}/4}
	+ (\epsilon_{+}\rightarrow \epsilon_{-})
	\biggr)
\biggr], \label{suppl-eq:coeff1} \\
c_{2}(\theta)
&=\frac{\mu}{5\epsilon+3\mu}
\biggl[
	-2\frac{\epsilon^{2}+3\mu \epsilon+\mu^{2}}{\mu\epsilon}
	+\frac{2\epsilon+\mu}{\epsilon}
	\biggl(
		\frac{2\epsilon+\mu}{\epsilon_{+}/2+\mu}+ (\epsilon_{+}\to \epsilon_{-})
	\biggr)
\biggr],\label{suppl-eq:coeff2}
\end{align}
\end{subequations}
which are Eqs.~\eqref{eq:coeff1} and \eqref{eq:coeff2} in the main text.

To get back the original parameters, let us introduce
\begin{align}
R_{k}(t)=\varepsilon r_{k}(\varepsilon^{2}t)e^{-i\varepsilon^{2}\tilde{\omega}_{k}t}.
\label{suppl-eq:def-Rk}
\end{align}
Then, the solvability condition~\eqref{suppl-eq:solvability-condition} turns into
\begin{align}
i\dif{}{t}R_{k}(t)=\Delta R_{k}(t)-i\alpha R^{\ast}_{k}(t)
+\lambda\Bigl(|R_{k}(t)|^{2}R_{k}(t)+c_{1}(\theta)|R_{p}(t)|^{2}R_{k}(t)+c_{2}(\theta)R_{p}(t)^{2}R^{\ast}_{k}(t)\Bigr),
\end{align}
with
\begin{align}
\Delta=\biggl(\frac{\omega}{2}-E\biggr),\qquad \alpha=\mu\frac{A}{2}\frac{\epsilon}{E},\qquad \lambda=\mu\frac{5\epsilon+3\mu}{E},
\label{suppl-eq:def-coeffs}
\end{align}
which is the amplitude equation~\eqref{eq:amp-eq} in the main text.
Also, $\varepsilon\phi^{k/p}_{1}(t,\tau)$ with Eq.~\eqref{suppl-eq:Bogoliubov-basis} expressed in terms of $R_{k}(t)$ using Eq.~\eqref{suppl-eq:def-Rk} corresponds to Eq.~\eqref{eq:Bogoliubov-basis} in the main text.

\section{Derivation of the amplitude equation~\eqref{eq:amp-eq-with-Rkp}}
We again start with the GP equation~\eqref{suppl-eq:GPequation}.
Considering the growth of the mode with the wavevector $\bk+\bp$ associated with the growth of the $\bk$- and $\bp$-modes, we take the following order counting with respect to the small bookkeeping parameter $\varepsilon$:
\begin{align}
A=\varepsilon^{2}\tilA,\qquad
 \frac{\omega}{2}-E_{\bk}=\varepsilon \tilde{\omega}_{k},\qquad
 \frac{\omega}{2}-E_{\bp}=\varepsilon \tilde{\omega}_{p},\qquad
 \omega-E_{+}=\varepsilon \tilde{\omega}_{+}.
\end{align}
Introducing two slow timescales $\tau_{1}=\varepsilon t$ and $\tau_{2}=\varepsilon^{2}t$, we expand the wave  function as
\begin{align}
\Psi(t,\bx)
=\Psi_{\uni}(t)\Biggl[1+\sum_{n=1}^{\infty}\varepsilon^{n}\phi_{n}(t,\tau,\bx)\Biggr],
\end{align}
where we regard $\phi_{n}$ as a function not only of $t$ and $\bx$ but also of $\tau_{1}$ and $\tau_{2}$, and labeled the two slow times collectively as $\tau=(\tau_{1},\tau_{2})$ for brevity.
In this section, we suppose $\phi_{1}(t,\tau,\bx)$ to have the form of
\begin{align}
\phi_{1}(t,\tau,\bx)
=\phi^{k}_{1}(t,\tau)\cos(\bk\cdot\bx)
+ \phi^{p}_{1}(t,\tau)\cos(\bp\cdot\bx)
+ \phi^{+}_{1}(t,\tau)\cos((\bk+\bp)\cdot\bx).
\label{suppl-eq:phi1-with-Rkp}
\end{align}
Substituting the above reparameterizations and the expansion, together with $\partial_{t}\to \partial_{t}+\varepsilon\partial_{\tau_{1}}+\varepsilon^{2}\partial_{\tau_{2}}$, into Eq.~\eqref{suppl-eq:GPequation}, we find
\begin{subequations}
\begin{align}
&i\pd{}{t}\phi_{1}(t,\tau,\bx)+\frac{\nabla^{2}}{2m}\phi_{1}(t,\tau,\bx)-2\mu \Re[\phi_{1}(t,\tau,\bx)]=0, 
\label{suppl-eq:1st-eq-with-Rkp} \\
&i\pd{}{t}\phi_{2}(t,\tau,\bx)+\frac{\nabla^{2}}{2m}\phi_{2}(t,\tau,\bx)-2\mu \Re[\phi_{2}(t,\tau,\bx)]
=
-i\pd{}{\tau_{1}}\phi_{1}(t,\tau,\bx)
+\mu\Bigl(
	|\phi_{1}(t,\tau,\bx)|^{2}
	+2\Re[\phi_{1}(t,\tau,\bx)]\phi_{1}(t,\tau,\bx)
\Bigr), \label{suppl-eq:2nd-eq-with-Rkp} \\
&i\pd{}{t}\phi_{3}(t,\tau,\bx)+\frac{\nabla^{2}}{2m}\phi_{3}(t,\tau,\bx)-2\mu \Re[\phi_{3}(t,\tau,\bx)]
=
- i\pd{}{\tau_{1}}\phi_{2}(t,\tau,\bx)
- i\pd{}{\tau_{2}}\phi_{1}(t,\tau,\bx) \nonumber \\
&
+ \mu\Bigl(
	-2\tilA\sin(\omega t)\Re[\phi_{1}(t,\tau,\bx)]
	+|\phi_{1}(t,\tau,\bx)|^{2}\phi_{1}(t,\tau,\bx)
	+2\calK(\phi_{1}(t,\tau,\bx),\phi_{2}(t,\tau,\bx))
\Bigr), \label{suppl-eq:3rd-eq-with-Rkp}
\end{align}
\end{subequations}
for increasing orders of $\varepsilon$.
The solution of Eq.~\eqref{suppl-eq:1st-eq-with-Rkp} again describes the usual Bogoliubov quasiparticles and, with the use of Eq.~\eqref{suppl-eq:phi1-with-Rkp}, is found as
\begin{align}
\phi_{1}^{+}(t,\tau)
&= \biggl(1-\frac{\epsilon_{+}+2\mu}{E_{+}}\biggr)r_{+}(\tau)e^{iE_{+}t}
+ \biggl(1+\frac{\epsilon_{+}+2\mu}{E_{+}}\biggr)r^{\ast}_{+}(\tau)e^{-iE_{+}t},
\label{suppl-eq:Bogoliubov-basis-Rkp}
\end{align}
and the same as before for $\phi_{1}^{k/p}(t,\tau)$, except that $\tau$ represents the two variables $\tau_{1}$ and $\tau_{2}$.

Our task, as in the previous section, is to find the solvability conditions that Eqs.~\eqref{suppl-eq:2nd-eq-with-Rkp} and \eqref{suppl-eq:3rd-eq-with-Rkp} have no secular terms, although it is complicated by the presence of the new timescale $\tau_{1}$ and the additional mode with the wavevector $\bk+\bp$.
In the following, we consider only the minimal contributions necessary to remove the divergence from the previous amplitude equation as interactions between the $\bk+\bp$ mode and the $\bk$- or $\bp$-mode.

\subsubsection{Solvability conditions from the second-order equation}
For $\phi_{2}(t,\tau,\bx)$, it is sufficient to consider only modes resulting from scattering of the two modes in $\phi_{1}(t,\tau)$, as in the previous model.
This allows us to take $\phi_{2}(t,\tau,\bx)$ in the form of
\begin{align}
\phi_{2}(t,\tau,\bx)
&=
\phi_{20}(t,\tau)
+\phi^{k}_{22}(t,\tau)\cos(2\bk\cdot\bx)
+\phi^{p}_{22}(t,\tau)\cos(2\bp\cdot\bx)
+\phi^{+}_{22}(t,\tau)\cos(2(\bk+\bp)\cdot\bx)
 \nonumber \\
&\quad
+\phi^{+}_{2}(t,\tau)\cos((\bk+\bp)\cdot\bx)
+\phi^{-}_{2}(t,\tau)\cos((\bk-\bp)\cdot\bx)
+\phi^{k}_{21}(t,\tau)\cos(\bk\cdot\bx)
+\phi^{p}_{21}(t,\tau)\cos(\bp\cdot\bx).
\label{suppl-eq:expansion-phi2-Rkp}
\end{align}
Here, we dropped the contributions of the modes with the wavevectors $2\bk+\bp$ and $\bk+2\bp$ because they are irrelevant to the divergence in the previous model.

Plugging the expansion~\eqref{suppl-eq:expansion-phi2-Rkp} into Eq.~\eqref{suppl-eq:2nd-eq-with-Rkp}, one obtains the differential equation for each mode.
We first consider the differential equation for $\phi^{k}_{21}(t,\tau)$:
\begin{align}
i\pd{}{t}\phi^{k}_{21}(t,\tau)
-\epsilon_{\bk}\phi^{k}_{21}(t,\tau)
-2\mu\Re[\phi^{k}_{21}(t,\tau)]
=f_{21}^{k}(t,\tau)
\end{align}
with
\begin{align}
f^{k}_{21}(t,\tau)
&= -i\pd{}{\tau_{1}}\phi_{1}^{k}(t,\tau)
+\mu \calK(\phi^{p}_{1}(t,\tau),\phi^{+}_{1}(t,\tau)).
\end{align}
Taking the real and imaginary parts of the differential equation and diagonalizing them, we get
\begin{align}
\pd{}{t}\chi^{k}_{21}(t,\tau)-iE_{\bk}\chi^{k}_{21}(t,\tau)=g^{k}_{21}(t,\tau)
\label{suppl-eq:diagonalized-eq-21k}
\end{align}
and its complex conjugate with
\begin{align}
\chi^{k}_{21}(t,\tau)
=\frac{1}{\sqrt{2}}
\biggl(
	\frac{\epsilon_{\bk}+2\mu}{E_{\bk}}\Re[\phi^{k}_{21}(t,\tau)]
	-i\Im[\phi^{k}_{21}(t,\tau)]
\biggr), \\
g^{k}_{21}(t,\tau)
=\frac{1}{\sqrt{2}}
\biggl(
	\frac{\epsilon_{\bk}+2\mu}{E_{\bk}}\Im[f^{k}_{21}(t,\tau)]
	+i\Re[f^{k}_{21}(t,\tau)]
\biggr).
\end{align}
Since Eq.~\eqref{suppl-eq:diagonalized-eq-21k} with $g^{k}_{21}(t,\tau)=0$ has an oscillating solution $e^{iE_{\bk}t}$, the condition for Eq.~\eqref{suppl-eq:diagonalized-eq-21k} to have no secular term is equivalent to that $g^{k}_{21}(t,\tau)$ has no term proportional to $e^{iE_{\bk}t}$.
The solvability condition is obtained as
\begin{align}
i\pd{}{\tau_{1}}r_{k}(\tau)
= - \beta(\bk,\bp)
r_{+}(\tau) r^{\ast}_{p}(\tau)
e^{-i(\tilde{\omega}_{+}-\tilde{\omega}_{k}-\tilde{\omega}_{p})\tau_{1}},
\label{suppl-eq:solvability-con-rk}
\end{align}
with
\begin{align}
\beta(\bk,\bp)
=
\frac{\mu}{2}\biggl[
	\frac{\epsilon_{+}+2\mu}{E_{+}}
	- \frac{\epsilon_{\bp}+2\mu}{E_{\bp}}
	+ \frac{\epsilon_{\bk}}{E_{\bk}}\biggl(
		3
		+ \frac{\epsilon_{\bp}+2\mu}{E_{\bp}} \frac{\epsilon_{+}+2\mu}{E_{+}}
	\biggr)
\biggr]
\label{suppl-eq:def-lambda-kp}
\end{align}
Under this solvability condition, $\phi^{k}_{21}(t,\tau)$ is found as
\begin{align}
\phi^{k}_{21}(t,\tau)
& =
\biggl(1-\frac{\epsilon_{\bk}+2\mu}{E_{\bk}}\biggr)\tilde{r}_{k}(\tau)e^{iE_{\bk}t}
+ \biggl(1+\frac{\epsilon_{\bk}+2\mu}{E_{\bk}}\biggr)\tilde{r}^{\ast}_{k}(\tau)e^{-iE_{\bk}t} \nonumber \\
&\quad
+ (\varphi^{k}_{1}-\varphi^{k}_{2})r_{p}(\tau)r_{+}(\tau)e^{i(E_{\bp}+E_{+})t}
+ (\varphi^{k}_{1}+\varphi^{k}_{2})r^{\ast}_{p}(\tau)r^{\ast}_{+}(\tau)e^{-i(E_{\bp}+E_{+})t},
\label{suppl-eq:sol-phi21k}
\end{align}
where $\tilde{r}_{k}(\tau)$ is an arbitrary complex amplitude.
Although the coefficient $\varphi^{k}_{1/2}$ can be found by substituting this solution into the differential equation, it is irrelevant to our purpose.
For $\phi^{p}_{21}(t,\tau)$, we can find the same solvability condition~\eqref{suppl-eq:solvability-con-rk} and the solution~\eqref{suppl-eq:sol-phi21k} with $\bk$ and $\bp$ replaced with each other.

We next consider the differential equation for $\phi^{+}_{2}(t,\tau)$:
\begin{align}
i\pd{}{t}\phi^{+}_{2}(t,\tau)
-\epsilon_{+}\phi^{+}_{2}(t,\tau)
-2\mu\Re[\phi^{+}_{2}(t,\tau)]
=f_{2}^{+}(t,\tau)
\end{align}
with
\begin{align}
f^{+}_{2}(t,\tau)
&= -i\pd{}{\tau_{1}}\phi_{1}^{+}(t,\tau)
+\mu\calK(\phi^{k}_{1}(t,\tau),\phi^{p}_{1}(t,\tau)).
\end{align}
Taking the real and imaginary parts of the differential equation and diagonalizing them, we obtain
\begin{align}
\pd{}{t}\chi^{+}_{2}(t,\tau)-iE_{+}\chi^{+}_{2}(t,\tau)=g^{+}_{2}(t,\tau)
\label{suppl-eq:diagonalized-eq-2plus}
\end{align}
and its complex conjugate with
\begin{align}
\chi^{+}_{2}(t,\tau)
=\frac{1}{\sqrt{2}}
\biggl(
	\frac{\epsilon_{+}+2\mu}{E_{+}}\Re[\phi^{+}_{2}(t,\tau)]
	-i\Im[\phi^{+}_{2}(t,\tau)]
\biggr), \\
g^{+}_{2}(t,\tau)
=\frac{1}{\sqrt{2}}
\biggl(
	\frac{\epsilon_{+}+2\mu}{E_{+}}\Im[f^{+}_{2}(t,\tau)]
	+i\Re[f^{+}_{2}(t,\tau)]
\biggr).
\end{align}
Since Eq.~\eqref{suppl-eq:diagonalized-eq-2plus} with $g^{+}_{2}(t,\tau)=0$ has an oscillating solution $e^{iE_{+}t}$, the condition for Eq.~\eqref{suppl-eq:diagonalized-eq-2plus} to have no secular term is equivalent to that $g^{+}_{2}(t,\tau)$ has no term proportional to $e^{iE_{+}t}$.
Using $e^{i(E_{\bk}+E_{\bp})t}=e^{iE_{+}t}e^{i(\tilde{\omega}_{+}-\tilde{\omega}_{k}-\tilde{\omega}_{p})\tau_{1}}$, we get the solvability condition as
\begin{align}
i\pd{}{\tau_{1}}r_{+}(\tau)
=- \beta_{+}(\bk,\bp) r_{k}(\tau)r_{p}(\tau)
e^{i(\tilde{\omega}_{+}-\tilde{\omega}_{k}-\tilde{\omega}_{p})\tau_{1}},
\label{suppl-eq:solvability-con-rplus}
\end{align}
with
\begin{align}
\beta_{+}(\bk,\bp)
=\frac{\mu}{2}
\biggl[
	\frac{\epsilon_{\bk}+2\mu}{E_{\bk}}
	+ \frac{\epsilon_{\bp}+2\mu}{E_{\bp}}
	+ \frac{\epsilon_{+}}{E_{+}}\biggl(
		3
		- \frac{\epsilon_{\bk}+2\mu}{E_{\bk}} \frac{\epsilon_{\bp}+2\mu}{E_{\bp}}
	\biggr)
\biggr].
\label{suppl-eq:def-lambdaplus-kp}
\end{align}
Under this solvability condition, $\phi^{+}_{2}(t,\tau)$ is found as
\begin{align}
\phi^{+}_{2}(t,\tau)
& =
\biggl(1-\frac{\epsilon_{+}+2\mu}{E_{+}}\biggr)\tilde{r}_{+}(\tau)e^{iE_{+}t}
+ \biggl(1+\frac{\epsilon_{+}+2\mu}{E_{+}}\biggr)\tilde{r}^{\ast}_{+}(\tau)e^{-iE_{+}t} \nonumber \\
&\quad
+ (\varphi^{+}_{2}-\varphi^{+}_{4})r_{k}(\tau)r^{\ast}_{p}(\tau)e^{i(E_{\bk}-E_{\bp})t}
+ (\varphi^{+}_{2}+\varphi^{+}_{4})r^{\ast}_{k}(\tau)r_{p}(\tau)e^{-i(E_{\bk}-E_{\bp})t},
\label{suppl-eq:sol-phi2plus}
\end{align}
where $\tilde{r}_{+}(\tau)$ is an arbitrary complex amplitude, and $\varphi^{+}_{2}$ and $\varphi^{+}_{4}$ are given as solutions of Eq.~\eqref{suppl-eq:coefficients-eq-phi2plus}. 
Comparing Eq.~\eqref{suppl-eq:sol-phi2plus} with the previous solution~\eqref{suppl-eq:phi2plus}, one can see that the divergent coefficients $\varphi^{+}_{1}$ and $\varphi^{+}_{3}$ are removed once the solvability condition~\eqref{suppl-eq:solvability-con-rplus} is imposed.
In other words, the solvability condition~\eqref{suppl-eq:solvability-con-rplus} provides the interaction between the $(\bk+\bp)$-mode and the $\bk$- and $\bp$-modes to remove the divergence in the previous model.

For the coefficient functions of the other modes in the expansion~\eqref{suppl-eq:expansion-phi2-Rkp}, only the particular solution is sufficient for our purposes and can be obtained as before.

\subsubsection{Solvability conditions from the third-order equation}
We expand $\phi_{3}(t,\tau,\bx)$ as
\begin{align}
\phi_{3}(t,\tau,\bx)
= \phi^{k}_{31}(t,\tau)\cos(\bk\cdot\bx)
+ \phi^{p}_{31}(t,\tau)\cos(\bp\cdot\bx)
+ \phi^{+}_{31}(t,\tau)\cos((\bk+\bp)\cdot\bx)
+ \cdots,
\end{align}
where the ellipsis represents terms with other irrelevant bases.
From the third-order differential equation~\eqref{suppl-eq:3rd-eq-with-Rkp}, the differential equations for $\phi^{k}_{31}(t,\tau)$ and $\phi^{+}_{31}(t,\tau)$ are found in the form of
\begin{subequations}
\begin{align}
&i\pd{}{t}\phi^{k}_{31}(t,\tau)-\epsilon_{\bk}\phi^{k}_{31}(t,\tau)-2\mu\Re[\phi^{k}_{31}(t,\tau)]=f^{k}_{31}(t,\tau), \\
&i\pd{}{t}\phi^{+}_{31}(t,\tau)-\epsilon_{+}\phi^{+}_{31}(t,\tau)-2\mu\Re[\phi^{+}_{31}(t,\tau)]=f^{+}_{31}(t,\tau),
\end{align}
\label{suppl-eq:decomposed-third-eq}
\end{subequations}
with
\begin{subequations}
\begin{align}
&f^{k}_{31}(t,\tau)
= -i\pd{}{\tau_{1}}\phi^{k}_{21}(t,\tau)-i\pd{}{\tau_{2}}\phi^{k}_{1}(t,\tau)
+\mu\biggl\{
	-2\tilA \sin(\omega t) \Re[\phi^{k}_{1}(t,\tau)]
	+\frac{3}{4}|\phi^{k}_{1}(t,\tau)|^{2}\phi^{k}_{1}(t,\tau) \nonumber \\
&
	+|\phi^{p}_{1}(t,\tau)|^{2}\phi^{k}_{1}(t,\tau)
	+\frac{1}{2}\phi^{k \ast}_{1}(t,\tau)\phi^{p}(t,\tau)^{2}
	+|\phi^{+}_{1}(t,\tau)|^{2}\phi^{k}_{1}(t,\tau)
	+\frac{1}{2}\phi^{k \ast}_{1}(t,\tau)\phi^{+}(t,\tau)^{2} \nonumber \\
&
	+\calK(\phi^{k}_{1}(t,\tau),2\phi_{20}(t,\tau)+\phi^{k}_{22}(t,\tau))
	+\calK(\phi^{p}_{1}(t,\tau),\phi_{2}^{+}(t,\tau)+\phi^{-}_{2}(t,\tau))
	+\calK(\phi^{+}_{1}(t,\tau),\phi^{p}_{21}(t,\tau))
\biggr\}, \\
&f^{+}_{31}(t,\tau)
= -i\pd{}{\tau_{1}}\phi^{+}_{2}(t,\tau)-i\pd{}{\tau_{2}}\phi^{+}_{1}(t,\tau)
+\mu\biggl\{
	-2\tilA \sin(\omega t)\Re[\phi^{+}_{1}(t,\tau)]
	+\frac{3}{4}|\phi^{+}_{1}(t,\tau)|^{2}\phi^{+}_{1}(t,\tau) \nonumber \\
&
	+|\phi^{k}_{1}(t,\tau)|^{2}\phi^{+}_{1}(t,\tau)
	+\frac{1}{2}\phi^{+ \ast}_{1}(t,\tau)\phi^{k}(t,\tau)^{2}
	+|\phi^{p}_{1}(t,\tau)|^{2}\phi^{+}_{1}(t,\tau)
	+\frac{1}{2}\phi^{+ \ast}_{1}(t,\tau)\phi^{p}(t,\tau)^{2} \nonumber \\
&
	+\calK(\phi^{+}_{1}(t,\tau),2\phi_{20}(t,\tau)+\phi^{+}_{22}(t,\tau))
	+\calK(\phi^{p}_{1}(t,\tau),\phi_{21}^{k}(t,\tau))
	+\calK(\phi^{k}_{1}(t,\tau),\phi_{21}^{p}(t,\tau))
\biggr\}.
\end{align}
\end{subequations}

To find the solvability conditions from these differential equations, we take their real and imaginary parts and diagonalize them as
\begin{subequations}
\begin{align}
\pd{}{t}\chi^{k}_{31}(t,\tau)-iE_{\bk}\chi^{k}_{31}(t,\tau)&=g^{k}_{31}(t,\tau),
\label{suppl-eq:diagonalized-eq-31k-Rkp} \\
\pd{}{t}\chi^{+}_{31}(t,\tau)-iE_{+}\chi^{+}_{31}(t,\tau)&=g^{+}_{31}(t,\tau), 
\label{suppl-eq:diagonalized-eq-31plus-Rkp}
\end{align}
\end{subequations}
and their complex conjugates.
Here, $\chi^{k/+}_{31}(t,\tau)$ and $g^{k/+}_{31}(t,\tau)$ are introduced as
\begin{subequations}
\begin{align}
\chi^{k}_{31}(t,\tau)
=\frac{1}{\sqrt{2}}
\biggl(
	\frac{\epsilon_{\bk}+2\mu}{E_{\bk}}\Re[\phi^{k}_{31}(t,\tau)]
	-i\Im[\phi^{k}_{31}(t,\tau)]
\biggr), \\
\chi^{+}_{31}(t,\tau)
=\frac{1}{\sqrt{2}}
\biggl(
	\frac{\epsilon_{+}+2\mu}{E_{+}}\Re[\phi^{+}_{31}(t,\tau)]
	-i\Im[\phi^{+}_{31}(t,\tau)]
\biggr),
\end{align}
\end{subequations}
and
\begin{subequations}
\begin{align}
g^{k}_{31}(t,\tau)
&=\frac{1}{\sqrt{2}}
\biggl(
	\frac{\epsilon_{\bk}+2\mu}{E_{\bk}}\Im[f^{k}_{31}(t,\tau)]
	+i\Re[f^{k}_{31}(t,\tau)]
\biggr), \\
g^{+}_{31}(t,\tau)
&=\frac{1}{\sqrt{2}}
\biggl(
	\frac{\epsilon_{+}+2\mu}{E_{+}}\Im[f^{+}_{31}(t,\tau)]
	+i\Re[f^{+}_{31}(t,\tau)]
\biggr).
\end{align}
\end{subequations}
Eqs.~\eqref{suppl-eq:diagonalized-eq-31k-Rkp} and \eqref{suppl-eq:diagonalized-eq-31plus-Rkp} with $g^{k}_{31}(t,\tau)=0$ and $g^{+}_{31}(t,\tau)=0$ have oscillating solutions $e^{iE_{\bk}t}$ and $e^{iE_{+}t}$, respectively.
Accordingly, the conditions for Eqs.~\eqref{suppl-eq:diagonalized-eq-31k-Rkp} and \eqref{suppl-eq:diagonalized-eq-31plus-Rkp} to have no secular terms is equivalent to that $g^{k}_{31}(t,\tau)$ and $g^{+}_{31}$ have no terms proportional to $e^{iE_{\bk}t}$ and $e^{iE_{+}t}$, respectively.
The solvability conditions are found as
\begin{subequations}
\begin{align}
&i\pd{}{\tau_{1}}\tilde{r}_{k}(\tau)+i\pd{}{\tau_{2}}r_{k}(\tau)
=
-i\mu\frac{\tilA}{2}\frac{\epsilon_{\bk}}{E_{\bk}} r^{\ast}_{k}(\tau)e^{i\tilde{\omega}_{k}\tau_{1}}
-\beta(\bk,\bp)
\Bigl(
	\tilde{r}_{+}(\tau)r_{p}^{\ast}(\tau)
	+ r_{+}(\tau)\tilde{r}_{p}^{\ast}(\tau)
\Bigr) e^{i(\tilde{\omega}_{k}+\tilde{\omega}_{p}-\tilde{\omega}_{+})\tau_{1}} \nonumber \\
&
+ \mu \frac{5\epsilon_{\bk}+3\mu}{E_{\bk}}
\Bigl(
	|r_{k}(\tau)|^{2}r_{k}(\tau)
	+\tilde{c}_{1}(\bk,\bp)|r_{p}(\tau)|^{2}r_{k}(\tau)
	+c_{2}(\bk,\bp)r_{p}(\tau)^{2}r^{\ast}_{k}(\tau)e^{2i(\tilde{\omega}_{k}-\tilde{\omega}_{p})\tau_{1}}
\Bigr), \\
&i\pd{}{\tau_{1}}\tilde{r}_{+}(\tau)+i\pd{}{\tau_{2}}r_{+}(\tau)
=
-\beta_{+}(\bk,\bp)
\Bigl(
	\tilde{r}_{k}(\tau)r_{p}(\tau)
	+ r_{k}(\tau)\tilde{r}_{p}(\tau)
\Bigr) e^{i(\tilde{\omega}_{+}-\tilde{\omega}_{k}-\tilde{\omega}_{p})\tau_{1}}
+ \mu \frac{5\epsilon_{+}+3\mu}{E_{+}} |r_{+}(\tau)|^{2}r_{+}(\tau).
\end{align}
\label{suppl-eq:solvability-con-third}
\end{subequations}
Here, we dropped the cubic interaction terms between the $(\bk+\bp)$-mode and the $\bk$- and $\bp$-modes, i.e., $|r_{+}(\tau)|^{2}r_{k/p}(\tau)$ and $|r_{k/p}(\tau)|^{2}r_{+}(\tau)$, as they are not essential to remove the divergence in the previous model.
The coefficients $\beta(\bk,\bp)$, $\beta_{+}(\bk,\bp)$ and $c_{2}(\bk,\bp)$ are given by Eqs.~\eqref{suppl-eq:def-lambda-kp}, \eqref{suppl-eq:def-lambdaplus-kp} and \eqref{suppl-eq:def-c2-kp}, respectively, and the coefficient $\tilde{c}_{1}(\bk,\bp)$ is given by
\begin{align}
\tilde{c}_{1}(\bk,\bp)&= 
\frac{E_{\bk}}{5\epsilon_{\bk}+3\mu}
\biggl[
	4\frac{\epsilon_{\bk}\epsilon_{\bp}-\mu^{2}}{\epsilon_{\bp}E_{\bk}}
	-\frac{1}{2}\varphi^{-}_{1}
	\biggl(-\frac{\epsilon_{\bp}+2\mu}{E_{\bp}}+3\frac{\epsilon_{\bk}}{E_{\bk}}\biggr)
	-\frac{1}{2}(\varphi^{+}_{2}+\varphi^{-}_{2})
	\biggl(\frac{\epsilon_{\bp}+2\mu}{E_{\bp}}+3\frac{\epsilon_{\bk}}{E_{\bk}}\biggr) \nonumber \\
&\quad
	-\frac{1}{2}\varphi^{-}_{3}
	\biggl(1+\frac{\epsilon_{\bp}+2\mu}{E_{\bp}}\frac{\epsilon_{\bk}}{E_{\bk}}\biggr)
	-\frac{1}{2}(\varphi^{+}_{4}+\varphi^{-}_{4})
	\biggl(1-\frac{\epsilon_{\bp}+2\mu}{E_{\bp}}\frac{\epsilon_{\bk}}{E_{\bk}}\biggr)
\biggr]. \label{suppl-eq:def-tilde-c1-kp}
\end{align}

\subsubsection{The resulting amplitude equation}
To combine the obtained solvability conditions and go back to the original parameters, we introduce
\begin{subequations}
\begin{align}
R_{k}(t)
&=\Bigl(
	\varepsilon r_{k}(\varepsilon t,\varepsilon^{2}t)
	+ \varepsilon^{2} \tilde{r}_{k}(\varepsilon t,\varepsilon^{2}t)
\Bigr)e^{-i\varepsilon \tilde{\omega}_{k}t}, \\
R_{p}(t)
&=\Bigl(
	\varepsilon r_{p}(\varepsilon t,\varepsilon^{2}t)
	+ \varepsilon^{2} \tilde{r}_{p}(\varepsilon t,\varepsilon^{2}t)
\Bigr)e^{-i\varepsilon \tilde{\omega}_{p}t}, \\
R_{+}(t)
&=\Bigl(
	\varepsilon r_{+}(\varepsilon t,\varepsilon^{2}t)
	+ \varepsilon^{2} \tilde{r}_{+}(\varepsilon t,\varepsilon^{2}t)
\Bigr)e^{-i\varepsilon \tilde{\omega}_{+}t}.
\end{align}
\end{subequations}
Then, with the use of the solvability conditions~\eqref{suppl-eq:solvability-con-rk}, \eqref{suppl-eq:solvability-con-rplus}, and \eqref{suppl-eq:solvability-con-third}, the time derivatives of these amplitudes are computed as
\begin{subequations}
\begin{align}
i\dif{}{t}R_{k}(t)
&= \Delta R_{k}(t)
- i\alpha R_{k}^{\ast}(t)
- \beta(\theta) R_{+}(t) R^{\ast}_{p}(t)
+\lambda\Bigl[
	|R_{k}(t)|^{2}R_{k}(t)
	+\tilde{c}_{1}(\theta)|R_{p}(t)|^{2}R_{k}(t)
	+c_{2}(\theta)R_{p}(t)^{2}R^{\ast}_{k}(t)
\Bigr], \\
i\dif{}{t}R_{+}(t)
&= \Delta_{+}(\theta) R_{k}(t)
- \beta_{+}(\theta) R_{k}(t) R_{p}(t)
+\lambda_{+}(\theta)|R_{+}(t)|^{2}R_{+}(t),
\end{align}
\end{subequations}
where we assumed the absolute values of $\bk$ and $\bp$ to be equal to focus on the angle between them.
Then, with the use of $\epsilon=\epsilon_{\bk}=\epsilon_{\bp}$ and $E=E_{\bk}=E_{\bp}$, the coefficients become functions of the angle $\theta=\angle(\bk,\bp)\in[0,\pi/2]$ given by
\begin{align}
\Delta_{+}(\theta)=\omega-E_{+},\quad
\beta(\theta)=\mu\biggl(\frac{\epsilon-\mu}{E}+\frac{\epsilon_{+}+2\mu}{E_{+}}\biggr),\quad
\beta_{+}(\theta)=\mu\biggl(\frac{\epsilon+2\mu}{E}+\frac{\epsilon_{+}}{E_{+}}\frac{\epsilon-\mu}{\epsilon}\biggr)
,\quad \lambda_{+}(\theta)=\mu\frac{5\epsilon_{+}+3\mu}{E_{+}},
\end{align}
and
\begin{align}
\tilde{c}_{1}(\theta)
&=\frac{\mu}{5\epsilon+3\mu}
\biggl[
	4\frac{\epsilon^{2}-\mu^{2}}{\mu\epsilon}
	-\frac{(2\epsilon-\mu)(\epsilon+2\mu)+(2\epsilon^{2}+\mu^{2})\epsilon_{-}/(2\epsilon)}{E^{2}-E^{2}_{-}/4}
	+\biggl(
		\frac{2\epsilon+\mu}{\epsilon}\frac{2\epsilon+\mu}{\epsilon_{+}/2+\mu}
		+ (\epsilon_{+}\rightarrow \epsilon_{-})
	\biggr)
\biggr].
\end{align}
The other coefficients $\Delta,\alpha,\lambda$, and $c_{2}(\theta)$ are the same as before and given by Eqs.~\eqref{suppl-eq:def-coeffs} and \eqref{suppl-eq:coeff2}.
The coefficient $\tilde{c}_{1}(\theta)$ is the coefficient $c_{1}(\theta)$ with the divergent term at the angle satisfying $2E=E_{+}$ removed, and thus all coefficients have no divergence in $\theta\in[0,\pi/2]$.
Here, $2E=E_-$ cannot be satisfied for $\theta\in[0,\pi/2]$ and therefore is not resonant; note that it is sufficient to consider only the case with $\theta\in[0,\pi/2]$ because the case with $\theta\in[\pi/2,\pi]$ can be taken into account with the replacement $\theta\in[0,\pi/2]$ by $\pi-\theta$.

Finally, assuming that $\bk$ and $\bp$ satisfy the resonance condition $\Delta=0$, and including phenomenological dissipation coefficients $\Gamma$ and $\Gamma_{+}$, we arrive at
\begin{subequations}
\begin{align}
i\dif{}{t}R_{k}(t)
&= -i\Gamma R_{k}(t)
- i\alpha R_{k}^{\ast}(t)
- \beta(\theta) R_{+}(t) R^{\ast}_{p}(t)
+\lambda\Bigl[
	|R_{k}(t)|^{2}R_{k}(t)
	+\tilde{c}_{1}(\theta)|R_{p}(t)|^{2}R_{k}(t)
	+c_{2}(\theta)R_{p}(t)^{2}R^{\ast}_{k}(t)
\Bigr], \\
i\dif{}{t}R_{+}(t)
&= -i\Gamma_{+}R_{k}(t)
+\Delta_{+}(\theta) R_{k}(t)
- \beta_{+}(\theta) R_{k}(t) R_{p}(t)
+\lambda_{+}(\theta)|R_{+}(t)|^{2}R_{+}(t),
\end{align}
\end{subequations}
with $\Delta_{+}(\theta)=2E-E_{+}$, which are the amplitude equations of Eq.~\eqref{eq:amp-eq-with-Rkp} in the main text.

\section{The stability to changes in the grid angle}
We discuss the stability to changes in the grid angle based on the three-mode model of Eq.~\eqref{eq:amp-eq-with-Rkp} in the main text, and show that the grid with $\theta=\pi/2$ is the most stable.
To take into account multiple superimposed grids, it is necessary to simultaneously consider three or more excitations satisfying the first resonance condition.
Let us take $N$ excited modes satisfying the first resonance condition and denote their wavevectors as $\bk_{m}$ for $m=1,2,\ldots, N$.
Since the first resonance condition $\omega/2=E_{\bk_{m}}$ fixes the absolute values of the wavevectors, $\bk_m$ is specified only by the angle $\theta_{m}\in[0,\pi]$ from a given reference direction, and we denote the corresponding amplitude as $R(t,\theta_{m})$.
This amplitude $R(t,\theta_{m})$ can be regarded as an angle-by-angle discretized distribution function of the excitations.

Since the pairwise interaction between excitations is already obtained as the amplitude equation~\eqref{eq:amp-eq-with-Rkp}, the time-evolution equation for this distribution $R(t,\theta_{m})$  can be found as
\begin{subequations}
\begin{align}
&i\dif{}{t}R(t,\theta_{m})
= 
-i\Gamma R(t,\theta_{m})
-i\alpha R^{\ast}(t,\theta_{m})
-\sum_{n(\neq m)}\beta(|\theta_{m}-\theta_{n}|)R_{+}(t,\theta_{m},\theta_{n})R^{\ast}(t,\theta_{n}) \nonumber \\
&
+ \lambda
\Bigl[
	|R(t,\theta_{m})|^{2}R(t,\theta_{m})
	+\sum_{n(\neq m)}\tilde{c}_{1}(|\theta_{m}-\theta_{n}|)|R(t,\theta_{n})|^{2}R(t,\theta_{m})
	+\sum_{n(\neq m)}c_{2}(|\theta_{m}-\theta_{n}|)R(t,\theta_{n})^{2}R^{\ast}(t,\theta_{m})
\Bigr], \\
&i\dif{}{t}R_{+}(t,\theta_{m},\theta_{n})
= 
-i\Gamma_{+}R_{+}(t,\theta_{m},\theta_{n})
+\Delta(|\theta_{m}-\theta_{n}|)R_{+}(t,\theta_{m},\theta_{n}) \nonumber \\
&
-\beta_{+}(|\theta_{m}-\theta_{n}|)R(t,\theta_{m})R(t,\theta_{n})
+\lambda_{+}(|\theta_{m}-\theta_{n}|)|R_{+}(t,\theta_{m},\theta_{n})|^{2}R_{+}(t,\theta_{m},\theta_{n}),
\end{align}
\label{suppl-eq:distribution}
\end{subequations}
where $R_{+}(t,\theta_{m},\theta_{n})$ is the complex amplitude of the mode with $\bk_{m}+\bk_{n}$, and $\bk_{m}$ and $\bk_{n}$ are specified by $\theta_{m}$ and $\theta_{n}$, respectively.
Here, the angle-dependent coefficients in $\theta\in[\pi/2,\pi]$ are defined to be symmetric around $\theta=\pi/2$.
The three-mode model discussed in the main text corresponds to the $N=2$ case of this model.

\begin{figure}[hb]
 \centering
 \includegraphics[width=1.0\linewidth]{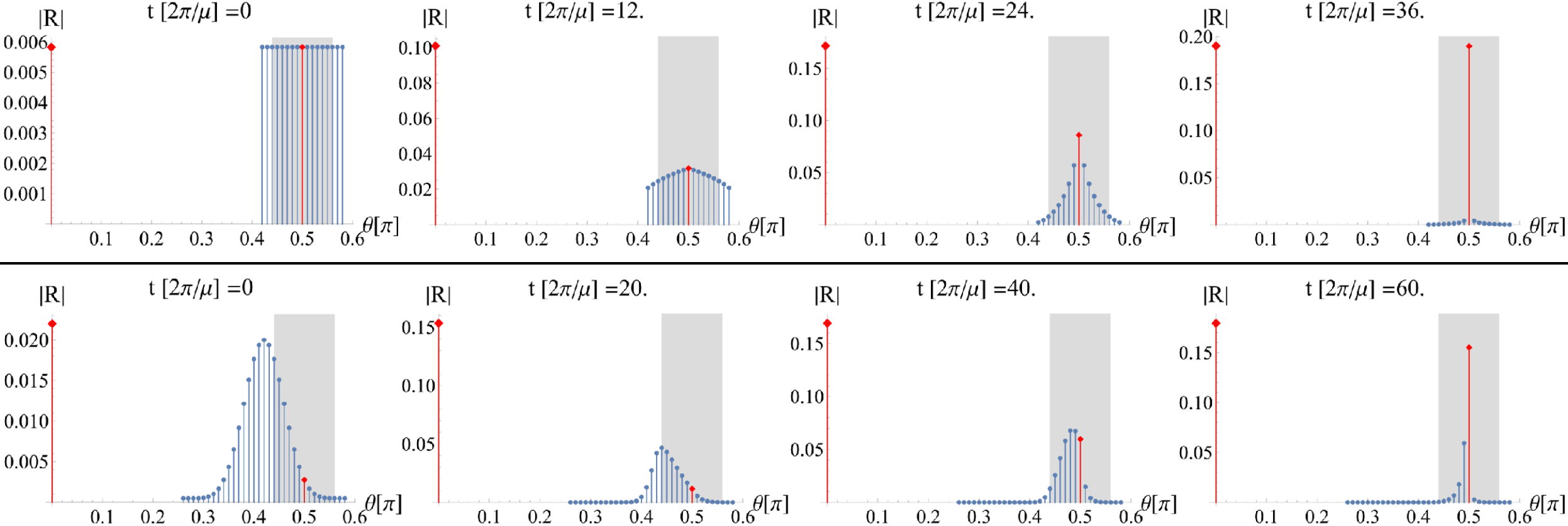}
 \caption{\label{suppl-fig:distribution}
Time evolution of the absolute value of the discretized angle-by-angle distribution function $|R(t,\theta_m)|$ from Eq.~\eqref{suppl-eq:distribution} with parameters $(A,\omega/\mu,\Gamma/\alpha,\Gamma_{+}/\alpha)=(0.6,0.5,1.0)$.
The gray shaded area represents the stable zone, i.e., the range of angles where the grid is stable with respect to a reference peak at $\theta=0$; the red lines represent $\theta=0$ and $\theta=\pi/2$.
(Top) The initial distribution is a uniform distribution around $\theta=\pi/2$.
(Bottom) The initial distribution is a broad peak with its center at $\theta=0.42\pi$ outside of the stable zone.
}
\end{figure}

By numerically simulating this model~\eqref{suppl-eq:distribution} for various initial conditions, we can confirm that excitations separated by the angle $\theta=\pi/2$ survive and grow most stably.
Figure~\ref{suppl-fig:distribution} plots the time evolution of the absolute value of the discretized distribution function, $|R(t,\theta_m)|$ from equation~\eqref{suppl-eq:distribution}, for a choice of two initial conditions; (top) a uniform distribution in the stable regime, and (bottom) a broad peak with its center in the unstable region.
In both cases, there is an additional peak at $\theta=0$ as a reference, and the peaks at $\theta=0$ and $\theta=\pi/2$ eventually survive and grow.

\end{document}